\documentclass[11pt]{article}

\usepackage{amsthm}
\usepackage{amsmath}
\usepackage{natbib}
\usepackage{graphicx}
\usepackage{authblk}

\usepackage{wasysym}

\usepackage{placeins}

\usepackage[
  a4paper,
  left=2cm,
  right=2cm,
  top=2.5cm,
  bottom=3cm
]{geometry}

\newcommand{\savg}{\bar{D}_\text{ice}}
\newcommand{\Dmono}{\bar{D}_\text{mono}}
\newcommand{\Dnon}{\hat{D}_\text{max}}
\newcommand{\Dmax}{{D}_\text{max}}
\newcommand{\Dmin}{{D}_\text{min}}
\newcommand{\Dnorm}{{D}_\text{norm}}
\newcommand{\MR}{\mathit{MR}}
\newcommand{\mr}{\mathit{MR}}
\newcommand{\sigmaD}{\sigma}
\newcommand{\half}{\frac{1}{2}}
\newcommand{\flake}{{\scalebox{0.6}{\hexstar}}}  
\newcommand{\plate}{{\scalebox{0.6}{\hexagon}}}  
\newcommand{\halfD}{_{\frac{1}{2}}^\flake}
\newcommand{\halfP}{_{\frac{1}{2}}^\plate}

\newcommand{\McSnow}{{\it McSnow}}
\newcommand{\mrime}{m_r}
\newcommand{\vrime}{V_r}
\newcommand{\mliq}{m_\ell}
\newcommand{\mfrz}{m_f}
\newcommand{\vice}{V_i}

\title{On the geometry of aggregate snowflakes}

\author{
  Axel Seifert\thanks{Corresponding author: axel.seifert@dwd.de} \\
  \vspace{-3mm}
  {\small Deutscher Wetterdienst, Offenbach, Germany}
  \and
  \vspace{-3mm}
  Fabian Jakub\\
  \vspace{-3mm}
  {\small Deutscher Wetterdienst, Offenbach, Germany}
  \and
  \vspace{-3mm}
  Christoph Siewert\\
  \vspace{-3mm}
  {\small Deutscher Wetterdienst, Offenbach, Germany}
  \and
  \vspace{-3mm}
  Leonie von Terzi\\
  \vspace{-3mm}
  {\small Ludwig-Maximilians University of Munich, Munich, Germany}
  \and
  \vspace{-3mm}
  Stefan Kneifel\\
  \vspace{-3mm}
  {\small Ludwig-Maximilians University of Munich, Munich, Germany}
  \vspace{-5mm}
}

\date{} 

\begin{document}

\maketitle


\begin{abstract}
Snowflakes play a crucial role in weather and climate. A significant
portion of precipitation that reaches the surface originates as ice,
even when it ultimately falls as rain. Contrary to the popular image
of symmetric, dendritic crystals, most large snowflakes are irregular
aggregates formed through the collision of primary ice crystals, such
as hexagonal plates, columns, and dendrites. These aggregates exhibit
complex, fractal-like structures, particularly at large sizes. Despite
this structural complexity, each aggregate snowflake is unique, with
properties that vary significantly around the mean — variability that
is typically neglected in weather and climate models.  Using a
physically based aggregation model, we generate millions of synthetic
snowflakes to investigate their geometric properties. The resulting
dataset reveals that, for a given monomer number (cluster size) and
mass, the maximum dimension follows approximately a lognormal
distribution. We present a parameterization of aggregate geometry that
captures key statistical properties, including maximum dimension,
aspect ratio, cross-sectional area, and their joint correlations. This
formulation enables a stochastic representation of aggregate
snowflakes in Lagrangian particle models. Incorporating this
variability improves the realism of simulated fall velocities,
enhances growth rates by aggregation, and broadens Doppler radar
spectra in closer agreement with observations.

\end{abstract}

\section{Introduction}

Snowflakes have fascinated scientists for centuries, dating back to
the 17th century when German astronomer Johannes Kepler first
speculated on the physics behind their hexagonal shapes
\citep{Ball-2011}. He proposed that the hexagonal geometry of
snowflakes might be explained by a close packing of spheres — an idea
that led to what is now known as Kepler’s conjecture
\citep{Hales-2024}. Today, we understand that the intricate shape of
dendritic snow crystals arises from the hexagonal molecular structure
of ice, combined with the Mullins–Sekerka instability that promotes
the growth of filament-like branches \citep{Libbrecht-2019}.  However,
most snowflakes observed in the atmosphere are not pristine dendritic
crystals but rather aggregates of primary ice crystals formed through
collisions. These primary ice crystals — or monomers — may be
hexagonal plates, columns, or dendrites. The resulting aggregate
snowflakes are structurally complex and exhibit fractal scaling
behavior when composed of many monomers, particularly at large cluster
sizes \citep{Westbrook-2004a,Westbrook-2004b}.  In weather and climate
models, and even in the most detailed cloud-resolving models,
snowflake geometry is usually prescribed using empirical relations and
simple power-law fits
\citep[e.g.,][]{Mitchell-1990,Mitchell-1996}. Yet measurements of the
mass–size relationships for aggregates consistently show substantial
scatter around the mean. This variability is evident in both classic
studies \citep{Locatelli-Hobbs-1974,Mitchell-1990} and more recent
measurements using high-resolution imaging and machine learning
\citep{Leinonen-2021}. Importantly, this scatter cannot be fully
explained by measurement error alone. To a large extent, it is a
reflection of natural variability in aggregate snowflake geometry. For
a given mass and monomer number, aggregate snowflakes can vary from
compact and nearly spherical to chain-like and highly elongated. This
geometric diversity directly affects particle properties such as
terminal fall velocity: more compact aggregates fall faster than
elongated ones of the same mass. Since snowflake growth via
aggregation is largely driven by differential sedimentation,
variations in fall velocity can enhance the growth rate. This effect
is similar to atmospheric turbulence leading to an increase in the
collision rate by introducing variability in the particle velocity
\citep{Shaw-2003,Grabowski-Wang-2013,Onishi-Seifert-2016,Chellini-Kneifel-2024}.

Given these implications, it is important to characterize not only the
mean behavior of snowflakes but also the statistical variability of
their geometry. Specifically, we investigate the probability
distribution governing deviations in maximum dimension for a given
mass and monomer number. This leads to the development of a stochastic
model for aggregate snowflake geometry, suitable for use in Lagrangian
Monte Carlo particle simulations of cloud microphysics. Decades after
the pioneering work of \cite{Telford-1955} and \citet{Gillespie-1975}, Lagrangian
particle-based models have gained renewed interest and offer new
opportunities to address longstanding challenges in cloud physics
\citep{Shima-2009,Brdar-Seifert-2018,Morrison-2020,Shima-2020,Chandrakar-2021,Welss-2024,Morrison-2024}.

Previous studies of aggregate snowflake geometry, including those
by \citet{Dunnavan-2019}, \citet{Dunnavan-2021},
and \citet{Przybylo-2022}, have examined snowflake shape, density, and
fall speed using observational datasets and Monte Carlo aggregation
models. \citet{Dunnavan-2019} employed bivariate statistical methods
to show how monomer habit and aggregation history influence the
observed geometric variability. \citet{Przybylo-2022} expanded on this
by using the Ice Particle and Aggregate Simulator (IPAS) to generate a
large database of computer-simulated aggregates, aimed at informing
bulk microphysical parameterizations. \citet{Dunnavan-2021} focused on
how variability in snowflake geometry affects terminal fall velocity,
a process that is revisited in this study. While these studies offer
important insights, their statistical frameworks are not readily
applicable in Lagrangian particle models (LPMs). To address this, we
develop a stochastic parameterization tailored for use in LPMs. In
this approach, key geometric properties such as maximum dimension,
aspect ratio, and cross-sectional area are probabilistically sampled
based on three inputs: ice mass, monomer number, and monomer
habit. This allows for efficient and physically consistent generation
of snowflake geometries, supporting both LPMs and remote sensing
forward operators.

The paper is organized as follows: Section 2 introduces the
aggregation model and the dataset that underpins our
parameterization. Section 3 describes the statistical model of
snowflake geometry, focusing on maximum dimension, aspect ratio, and
projected area, derived as functions of mass, monomer number, and
habit. In Section 4, we demonstrate the implementation of this
parameterization within the Lagrangian particle model McSnow. In Section 5 we investigate the influence of the new parameterization on common radar variables, also introducing a new scattering database developed for this study. Section
6 concludes with a summary and outlook.


\section{Aggregation modeling}\label{sec:agg_modelling}

The aggregation model described by \citet{Leinonen-Szyrmer-2015}
and \citet{Leinonen-Moisseev-2015} constructs three-dimensional
representations of aggregate snowflakes by combining primary crystals
of predefined shapes such as hexagonal columns, plates, or
dendrites. This approach is conceptually similar to the model
presented by \citet{Westbrook-2004a,Westbrook-2004b}. Our analysis
focuses exclusively on unrimed snowflakes, although the geometry of
rimed aggregates has been addressed in previous
work \citep{Seifert-2019}. The same aggregation framework has also
been used to study the physical properties of snowflake
aggregates \citep{Karrer-2020}, their scattering
characteristics \citep{Ori-2021}, and to develop retrieval algorithms
based on in-situ observations \citep{Leinonen-2021}. In addition, the
model has been applied to generate 3D-printed snowflake analogs for
laboratory experiments \citep{Kobschall-2023}.

Building on this framework, we refine the constant penetration depth
concept introduced by \citet{Leinonen-Moisseev-2015} to allow the
formation of slightly more compact aggregates.  The basic
implementation for attaching monomers and aggregates follows a
procedure similar to the game \textit{Tetris}.  Two particles are
randomly aligned in the vertical direction such that their horizontal,
axis-aligned bounding boxes overlap.  The lower one is then moved
upward until it intersects the upper one at the point of minimum
vertical distance. If no intersection occurs, the orientation is
randomized again until contact is achieved. This method, however,
tends to produce very fragile connections, comparable to two needles
just touching at their tips, resembling the near-touching fingers in
Michelangelo’s \textit{The Creation of Adam}.

To increase the compactness of the aggregates, the original approach
extended the contact depth by a fixed amount. Once a contact point was
found, the lower particle was moved an additional 80\,$\mu$m
upward. Although this adjustment led to denser structures on average,
it also introduced unphysical side effects, sometimes producing
disconnected or overly porous aggregates, particularly for small
particles.  In our revised approach, we aim to improve the realism of
the intersection process by excluding the outer portion of a
particle’s volume during contact evaluation. Specifically, we define a
spherical boundary that encloses 50\,\% of the particle’s material
volume and consider only the inner region for collision detection. For
needle-shaped monomers, this effectively removes the tips from
consideration and reduces the usable intersection area by about one
quarter. For plates, halving the usable volume corresponds to
excluding approximately $1 - 1/\sqrt{2} \approx 29\,\%$ of the
diameter. For a sphere, the excluded fraction becomes $1 -
1/\sqrt[3]{2} \approx 20\,\%$. In a more extreme case, such as a
compact ice sphere with a long protruding arm, the arm would be
excluded entirely from the intersection calculation. Finally, to avoid
excessive exclusion in large particles, we cap the masked region to a
maximum shell thickness of 120\,$\mu$m.

Primary crystals are drawn from a truncated exponential size
distribution, with a minimum size of $\savg/4$ and a maximum of
$100\savg$, where $\savg$ is the mean monomer size, ranging from
50\,$\mu$m to 500\,$\mu$m. The aggregation model operates at a finite
spatial resolution, which in our study is proportional to the mean
monomer size and defined as $\Delta = \savg / 20$. Monomer habits
include needles, plates, rosettes, and dendrites. For details on the
monomer geometries, see Figure~1 of \citet{Leinonen-Moisseev-2015}. In
this work, we restrict our attention to aggregates composed of
needles, plates, and dendrites.

To generate a comprehensive dataset, random samples are created with
monomer numbers $N$ ranging from 2 to 2048, and mean monomer sizes
$\savg$ from 50\,$\mu$m to 500\,$\mu$m. We produce five distinct
datasets: (1) needles only, (2) plates only, (3) dendrites only, (4)
mixtures of needles and plates, and (5) mixtures of needles and
dendrites. Following the terminology of \citet{Mitchell-1996}, we
refer to these as aggregates of mixtures, meaning that individual
aggregates contain monomers of different shapes. The monomer ratio
$\MR$ denotes the number fraction of needle monomers, with the
remainder being plates (or dendrites).

In our datasets, the
aggregates of mixtures combine prolate forms, such as needles, with
oblate ones, such as plates or dendrites. This choice is motivated by
habit predictions in Lagrangian particle models, in particular McSnow,
and also by the possibility that fragmentation introduces
needle-shaped monomers into regions where dendritic growth
occurs. While other combinations, like plate--dendrite mixtures, may
form in nature, they are not considered here. Each single-habit
dataset contains approximately one million samples, while the
mixed-habit datasets include about five million samples each to cover
a wide range of habit ratios.

For each aggregate, the database records several properties: the
number of monomers $N$, the mean monomer size $\savg$, the total mass
$m$, the maximum dimension $D$, the vertically projected
cross-sectional area $A$, and three characteristic lengths $L_x$,
$L_y$, and $L_z$, which represent the maximum extent of the aggregate along the
respective spatial axes. After every collision, a principal axis
transformation is performed. As a result, the ordering $L_x > L_y > L_z$
holds for nearly all particles. However, a few exceptions with $L_x <
L_y$ occur due to the geometry of the principal axis transformation. Although the dataset includes the principal axes, these are not used in the subsequent analysis.

The horizontal aspect ratio, defined as $\phi_{xy} = L_x / L_y$, is
relevant for estimating terminal fall velocity, whereas the vertical
aspect ratio, $\phi_{xz} = L_z / L_x$, is, for example, accessible
through video disdrometer observations. For consistency with previous
studies \citep{Seifert-2019,Karrer-2020}, the maximum dimension $\Dmax$ is
defined as the diameter of the smallest sphere that completely
encloses the aggregate. While $\Dmax$ and $L_x$ are typically very
similar, they are not identical.

\section{Snowflake geometry}

The fractal geometry of snowflakes is traditionally described using
power-law relationships for both the mass–size and mass–area
dependencies \citep{Locatelli-Hobbs-1974, Mitchell-1990,
  Mitchell-1996}. Figure~\ref{fig:mass-size} shows the mass–size
relations for aggregates of needles, aggregates of plates, and
aggregates of dendrites. For monomer numbers greater than 10, the
aggregate mass $m$ scales with the maximum dimension $\Dmax$ via a
fractal exponent $p$ as $m \propto \Dmax^p$.

In this study, we find exponents ranging from 2.03 for aggregates of
needles, 2.08 for aggregates of dendrites, to 2.12 for aggregates of
plates. The power law fit uses only samples with $\Dmax > 0.5$ mm and
$N>5$.  It is important to note that the power-law fit describes only
the mean behavior; a significant scatter around the mean is evident in
Figure~\ref{fig:mass-size}. In the following, we aim to understand and
quantify this variability in more detail.

\def\mywidth{5.5cm}
\begin{figure}[t]
  \centering
  \begin{minipage}{\mywidth}
    a) aggregates of needles \\[3mm]
    \includegraphics[width=\textwidth,viewport= 60 50 960 880,clip=]
                    {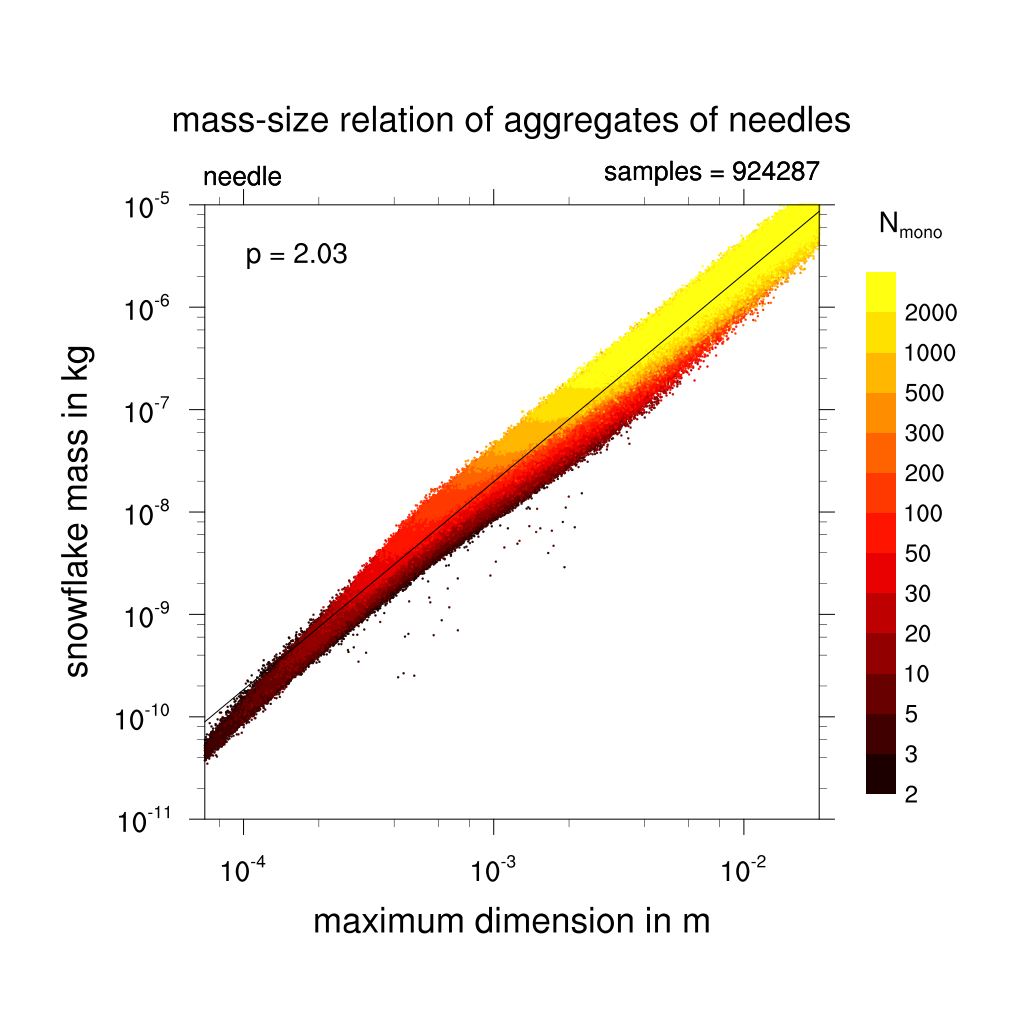}
  \end{minipage}
  \hfill
  \begin{minipage}{\mywidth}
    b) aggregates of plates \\[3mm]
    \includegraphics[width=\textwidth,viewport= 60 50 960 880,clip=]
                    {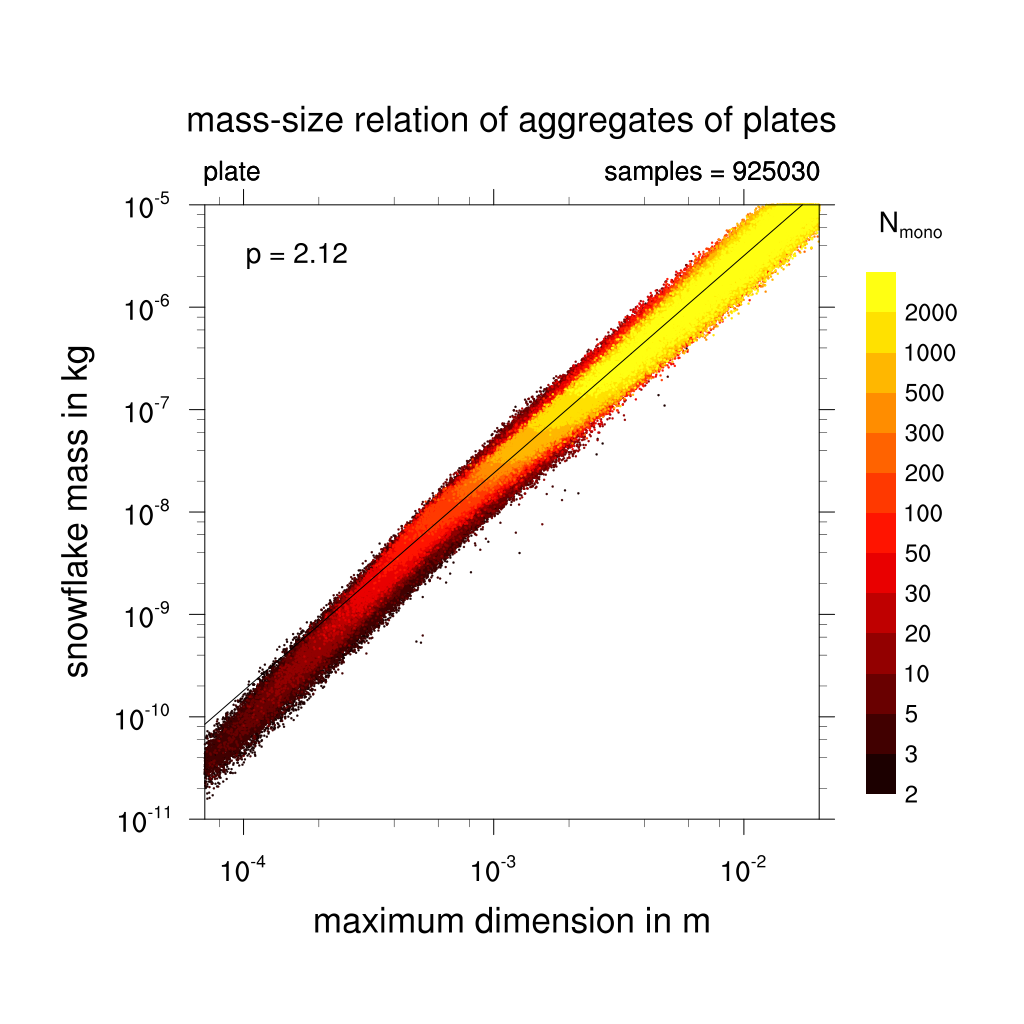}
  \end{minipage}
  \hfill
  \begin{minipage}{\mywidth}
    c) aggregates of dendrites \\[3mm]
    \includegraphics[width=\textwidth,viewport= 60 50 960 880,clip=]
                    {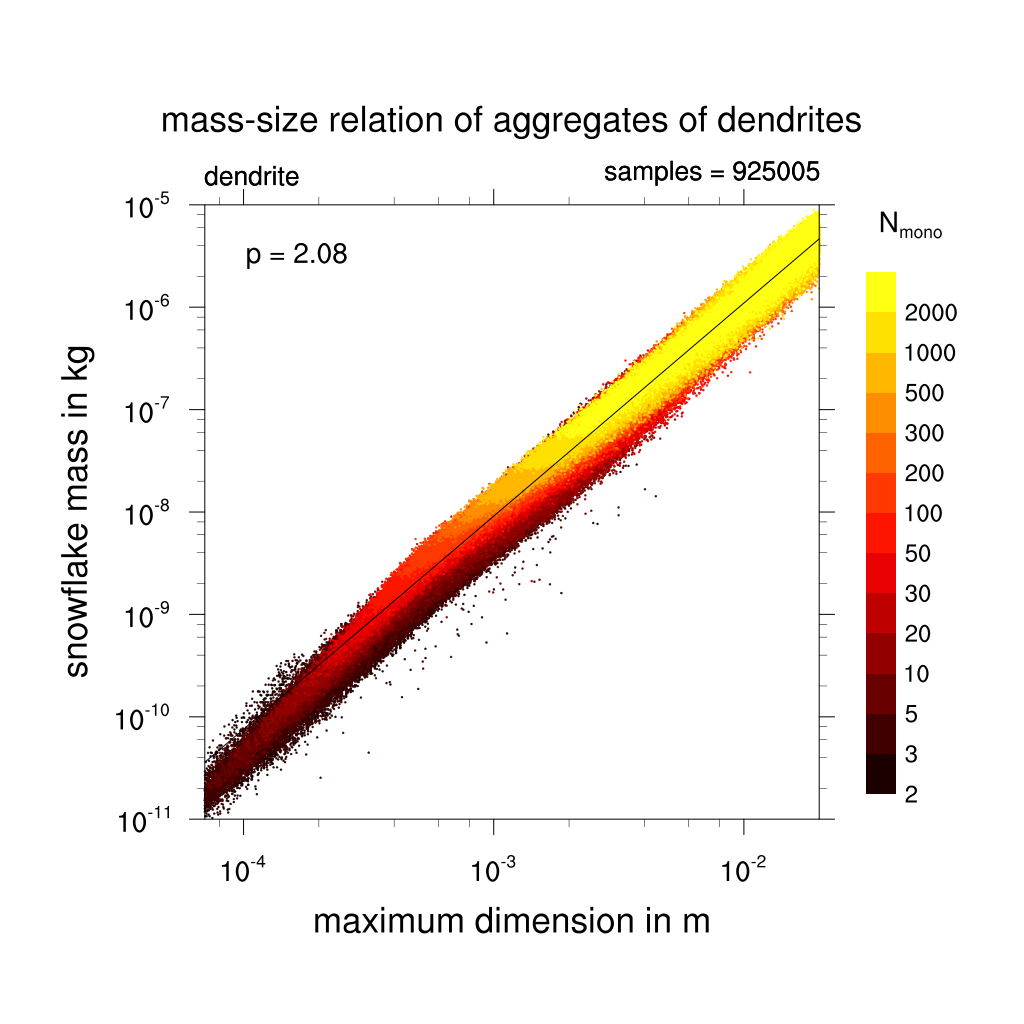}
  \end{minipage}
  \caption{Scatter plots showing the mass-size relations for aggregates of
  needles, plates, and dendrites. Colors indicate monomer number $N$,
  and the regression line represents the power-law fit with fractal
  exponent $p$ for $N>5$. 
  \label{fig:mass-size}}
\end{figure}

\subsection{Maximum dimension}

For a more detailed geometric model of aggregate snowflakes, the
maximum dimension is normalized using the mean monomer diameter,
denoted as $\Dmono$. This mean monomer diameter is derived from the
aggregate mass $m$ and the monomer number $N$ using
\begin{equation} \label{eq:Dmono}
  \Dmono = \left( \frac{m}{a_i \, N} \right)^{1/b_i},
\end{equation}
where $a_i$ and $b_i$ are the parameters of the mass–size relation for
the monomers (primary crystals). For needles these are $a_0=0.005$ and
$b_0=1.89$, for plates $a_1=0.788$ and $b_1=2.48$, and for dendrites
$a_2=0.013$ and $b_2=2.1$ with $m$ in kg and $\Dmono$ in m.

It is important to note that $\Dmono$ differs from $\savg$. The former
is specific to an individual aggregate, while the latter represents
the mean of the monomer size distribution for an ensemble of particles
as assumed in the aggregation model. With current observational
techniques, $\Dmono$ is generally not accessible via direct
measurements but can be readily computed in a Lagrangian particle
model.

The maximum dimension is made nondimensional using
\begin{equation} \label{eq:Dnon}
  \Dnon = \frac{\Dmax - \Dmin}{\Dmono},
\end{equation}
where $\Dmin = 15$ $\mu$m is a finite-size correction required due to
the truncated exponential distribution of the aggregation
model. Figure \ref{fig:Dnon-histograms} presents histograms of $\Dnon$
for various monomer numbers $N$. For a fixed $N$, the nondimensional
maximum dimension $\Dnon$ is approximately lognormally distributed.

To assess this approximation, statistical tests, including the
Kolmogorov–Smirnov (KS) test and the Anderson–Darling (AD) test, were
used to evaluate the goodness-of-fit to a lognormal distribution. Both
tests were applied to subsets of 500 samples for various monomer
numbers $N$. For all $N>2$, the KS test did not reject the null
hypothesis that the data follow a lognormal distribution at the 5 \%
significance level. However, the AD test, which is more sensitive to
deviations in the tails of the distribution, rejected the null
hypothesis for all values of $N$. These results suggest that while the
lognormal distribution provides a reasonable approximation, it may not
fully capture the true underlying distribution.

An alternative would be to apply the beta distribution, as in
\citet{Dunnavan-2019}, which has the advantage of having bounded
support, making it well-suited for modeling normalized geometric
quantities such as $\Dnorm$ or $\phi$ that are naturally restricted to
finite intervals. This can help avoid unphysical values and better
constrain the distribution's tails compared to the lognormal. Here, we
retain the lognormal distribution for its simplicity and computational
efficiency. In practice, generating lognormal random variates is
considerably faster than for the beta distribution: it requires only a
single normal variate and an exponential transform. In contrast,
sampling from the beta distribution often involves more complex
rejection sampling or transformations of gamma-distributed variables.
Given the high sample volume required in Lagrangian Monte Carlo
models, this performance difference can be substantial.

The approximately lognormal shape of the aggregate size distribution
observed in this study can be understood as a natural consequence of
stochastic growth processes with bounded inputs. Similar statistical
behavior has been reported in astrophysics, where the Stellar Initial
Mass Function (IMF) is often modeled as a lognormal distribution at
low and intermediate stellar masses, transitioning to a power-law tail
at higher masses \citep{Chabrier-2003,Basu-Jones-2004}. This form
arises from turbulent fragmentation and stochastic accretion, both of
which resemble bounded multiplicative processes where initial
conditions come from a limited distribution and growth proceeds via
variable-length aggregation constrained by physical limits. The
resulting IMF shows a lognormal core with deviations in the
tails. This pattern is mirrored in the synthetic data from our
snowflake aggregation model, where monomer sizes follow a truncated
exponential distribution and growth obeys a fractal scaling
law. Deviations from pure lognormality, especially in the tails, are
therefore not artifacts, but natural consequences of the underlying
physical constraints.


\def\mywidth{8.2cm}
\begin{figure}[t]
  \centering
  \begin{minipage}{\mywidth}
    a) aggregates of needles \\[3mm]
    \includegraphics[width=\textwidth,viewport= 70 170 550 590,clip=]
                    {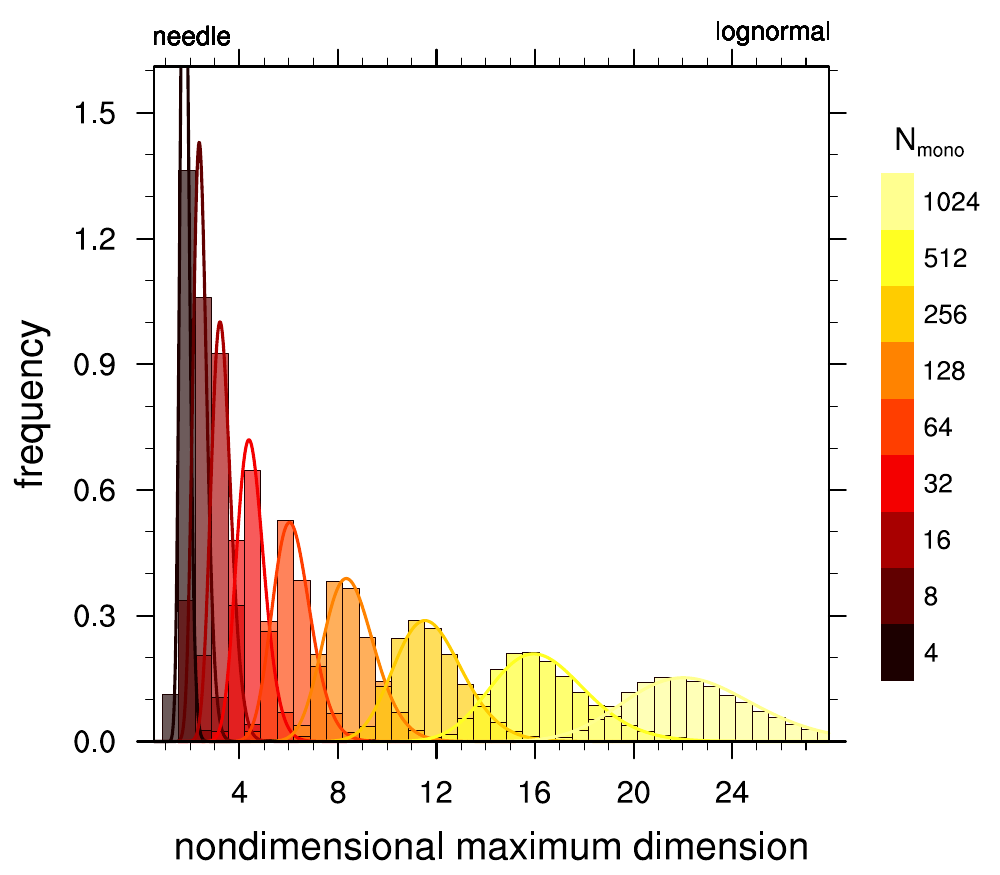}
  \end{minipage}
  \hfill
  \begin{minipage}{\mywidth}
    b) aggregates of plates \\[3mm]
    \includegraphics[width=\textwidth,viewport= 70 170 550 590,clip=]
                    {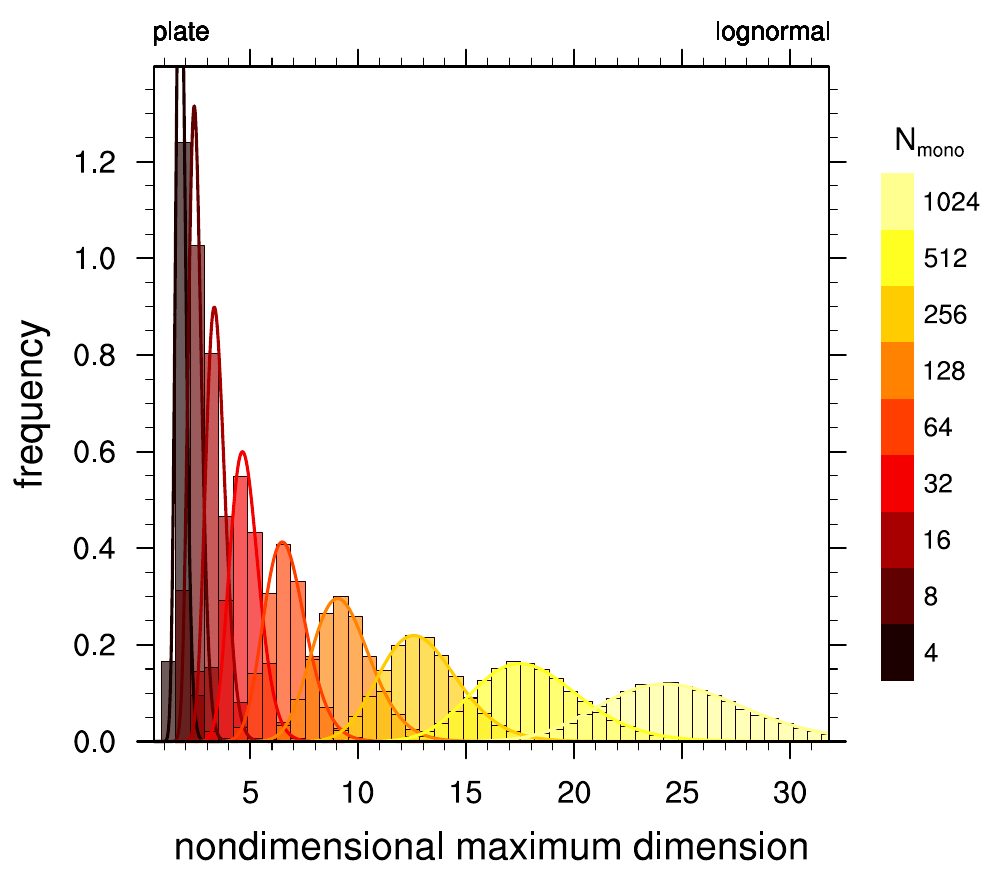}
  \end{minipage}
  \caption{Histograms of the nondimensional maximum dimension $\Dnon$ for
    aggregates of needles and aggregates of plates for different monomer numbers $N$.
    Solid lines represent lognormal
    distributions with the same mean and standard deviation as the
    data. The dendrite dataset looks similar and is not shown.
    \label{fig:Dnon-histograms}}
\end{figure}

Next, we parameterize the mean and standard deviation of these
lognormal distributions as functions of $N$. A power-law ansatz
suffices for the mean of $\Dnon$ consistent with a universal fractal
scaling of aggregates \citep{Westbrook-2004a,Westbrook-2004b}. This
leads us to define another normalized maximum dimension:
\begin{equation} \label{eq:Dnorm}
  \Dnorm = \Dnon \left ( \frac{N_0}{N} \right )^\eta = \frac{\Dmax - \Dmin}{\Dmono}  \left ( \frac{N_0}{N} \right )^\eta,
\end{equation}
such that $\Dnorm = 1$ corresponds to the mean aggregate snowflake,
independent of $N$. For $\eta$ we find 0.456 for aggregates of
needles, 0.474 for aggregates of plates, and 0.457 for aggregates of
dendrites. For aggregates of mixtures we find 0.459 for both the 50/50
mixture of needles and plates, and the 50/50 mixture of needles and
dendrites.  The corresponding values for $N_0$ are 1.136 for needles,
1.230 for plates, 1.161 for dendrites, 1.156 for the 50/50 mixture of
needles and plates, and 1.152 for the needles and dendrite
mixture. Note that with $\Dnorm=1$, Eq.~\eqref{eq:Dnorm} results in a
power-law for the mean mass-size relation.

\begin{figure}[p]
  \begin{flushleft}
    a) aggregates of needles with $N=64$ 
  \end{flushleft}
  \includegraphics[width=\textwidth,viewport= 0 1645 320 1730,clip=]{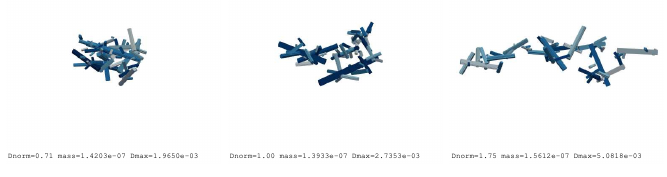}    
  \begin{flushleft}
    b) aggregates of plates with $N=64$
  \end{flushleft}
  \includegraphics[width=\textwidth,viewport= 0 1535 320 1620,clip=]{aggregates3d.pdf}
  \begin{flushleft}
    c) aggregates of dendrites with $N=64$ 
  \end{flushleft}
  \includegraphics[width=\textwidth,viewport= 0 1415 320 1500,clip=]{aggregates3d.pdf}
  \caption{\label{fig:small-agg-examples}
    Examples of aggregates of needles, aggregates of plates and aggregates of dendrites for monomer
    number $N=64$ and different $\Dnorm$. For each aggregate, the
    normalized diameter $\Dnorm$, the mass $m$ in kg, and the maximum
    dimension $\Dmax$ in m is given. The snowflakes with $\Dnorm < 1$
    (left column) are much smaller in terms of maximum dimension
    $\Dmax$ than the elongated snowflakes with $\Dmax > 1$ (right
    column). The central column shows an example of the mean
    aggregates snowflake with $\Dnorm=1$ for the given $N$. The
    aggregate mass $m$ in each row is similar, but not identical.}
\end{figure}

Snowflakes with $\Dnorm > 1$ tend to be more elongated and chain-like,
while those with $\Dnorm < 1$ are more compact and
spherical. Figures~\ref{fig:small-agg-examples} and
\ref{fig:large-agg-examples} illustrate examples of aggregates with
$\Dnorm < 1$, $\Dnorm \approx 1$, and $\Dnorm > 1$ for plate, needle and
dendrite monomers at different values of $N$. These visualizations
suggest an intuitive explanation for the lognormal shape of the
$\Dnorm$ distribution. Aggregates with $\Dnorm \approx 0.7$ are
already nearly spherical, with aspect ratios close to one, and thus
values much smaller than this are physically implausible. In contrast,
for $\Dnorm > 1$, values exceeding 1.5 are common, and the most
extreme, chain-like aggregates can reach up to $\Dnorm \approx
2$. This asymmetry, with bounded compactness on one side and a broader
range of elongation on the other, leads to a right-skewed
distribution, reflecting the underlying geometric constraints.

\begin{figure}[p]
  \begin{flushleft}
    a) aggregates of needles with $N=1024$ 
  \end{flushleft}
  \includegraphics[width=\textwidth,viewport= 0 605 320 685,clip=]{aggregates3d.pdf}
  \begin{flushleft}
    b) aggregates of plates with $N=1024$
  \end{flushleft}
  \includegraphics[width=\textwidth,viewport= 0 490 320 570,clip=]{aggregates3d.pdf}    
  \begin{flushleft}
    c) aggregates of dendrites with $N=1024$ 
  \end{flushleft}
  \includegraphics[width=\textwidth,viewport= 0 375 320 460,clip=]{aggregates3d.pdf}   
  \caption{\label{fig:large-agg-examples}
    As Figure \ref{fig:small-agg-examples}, but for a monomer number of $N=1024$.}
\end{figure}

For aggregates of needles we find that the standard deviation increases with $N$ as shown in Figure \ref{fig:std} 
and saturates for large $N$. The functional dependency can be parameterized with:
\begin{equation} \label{eq:sigma0}
  \sigmaD_0 = 0.121 - 0.032 \, \frac{2}{N}.
\end{equation}
For aggregates of plates the standard deviation shows no clear trend
but oscillates around
\begin{equation} \label{eq:sigma1}
  \sigmaD_1=0.142,  
\end{equation}
which is the mean of the data
for $N>10$ (Fig.~\ref{fig:std}a). For aggregate of dendrites in Fig.~\ref{fig:std}b we find that the standard deviation decreases slowly with $N$ following
\begin{equation} \label{eq:sigma2}
  \sigmaD_2 = 0.120 + 0.011 \left ( \frac{50}{N} \right )^{0.35}
\end{equation}
For aggregates of mixtures, the standard deviation is generally larger for small
$N$ compared to aggregates of only a single habit. This is because the
different monomer geometries allow for more variability and, hence,
higher standard deviations. The standard deviation then decreases
monotonically with $N$.  The following parameterization describes the
data for mixture aggregates of 50 \% needles and 50 \% plates (Fig.~\ref{fig:std}a):
\begin{equation}  \label{eq:sigma_mix1}
  \sigmaD\halfP = 0.120 + 0.016 \left ( \frac{50}{N} \right )^{0.4}
\end{equation}
and similarly for aggregates of 50 \% needles and 50 \% dendrites (Fig.~\ref{fig:std}b).
\begin{equation}  \label{eq:sigma_mix2}
  \sigmaD\halfD = 0.120 + 0.015 \left ( \frac{50}{N} \right )^{0.5}.
\end{equation}
Small aggregates with a mixture of needles and dendrites have the largest standard deviation in this dataset.

\def\mywidth{7.8cm}
\begin{figure}[t]
  \centering
  \begin{minipage}{\mywidth}
    a) aggregates of needles and plates \\[3mm]
    \includegraphics[width=\textwidth,viewport= 70 150 510 600,clip=]
                    {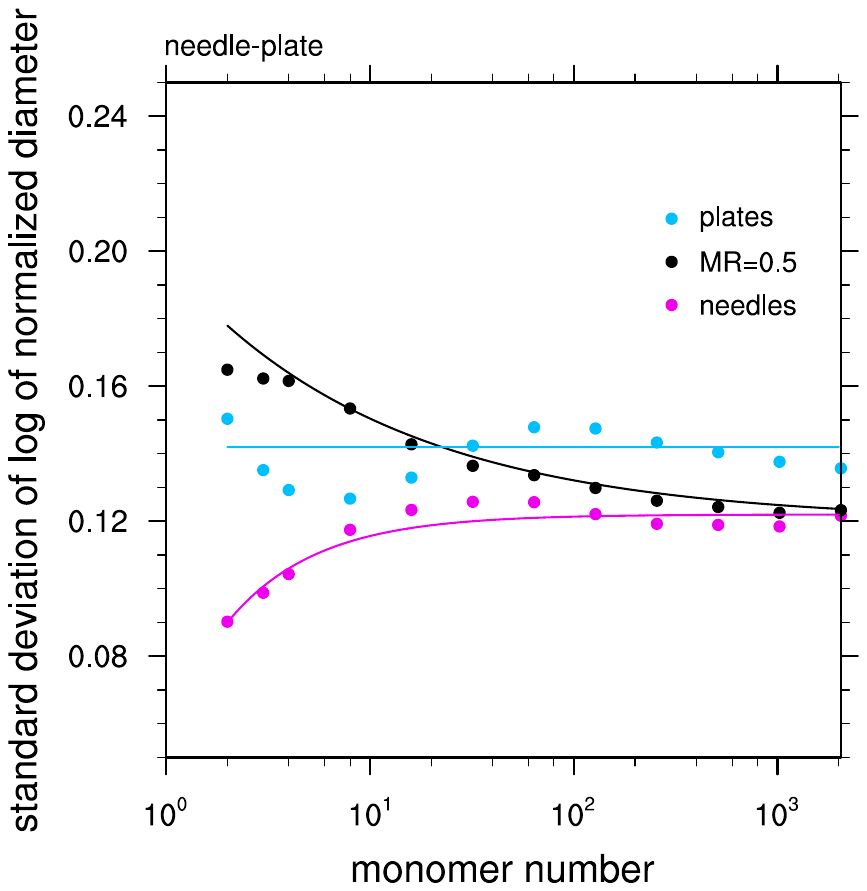}
  \end{minipage}
  \hfill
  \begin{minipage}{\mywidth}
    b) aggregates of needles and dendrites \\[3mm]
    \includegraphics[width=\textwidth,viewport= 70 150 510 600,clip=]
                    {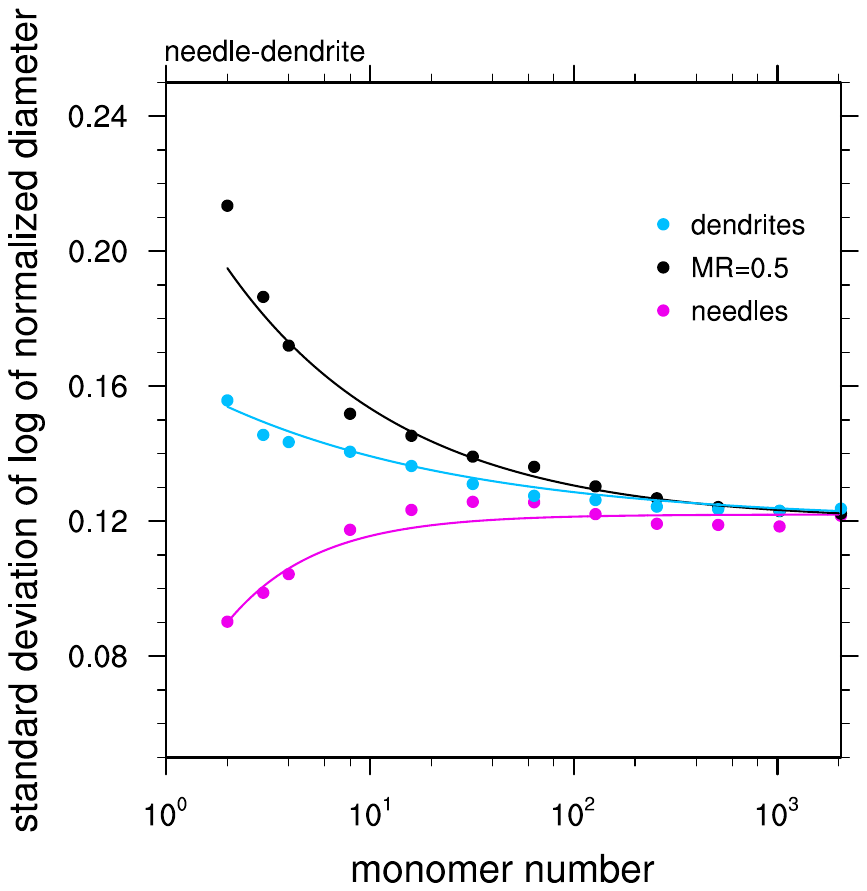}
  \end{minipage}
  \caption{Standard deviation of the normalized diameter $\Dnorm$ as a
    function of monomer number $N$ for aggregates of plates, aggregates
    of needles, and mixture aggregates of plates and needles (left), and
    aggregates of needles, aggregates of dendrites, and mixture aggregates
    of both (right). Aggregates of mixtures are shown for different
    monomer ratios $\MR$. The solid lines are the parameterizations
    using Eqs.~\eqref{eq:sigma0}--\eqref{eq:sigma_mix2}.
    \label{fig:std}}
\end{figure}

\subsection{Aspect ratio}
Aggregate snowflakes are inherently anisotropic; their aspect ratio
depends on both the monomer shape and monomer number $N$. In addition,
the aspect ratio $\phi$ is correlated with the normalized maximum
dimension $\Dnorm$, as can already be seen in
Figs.~\ref{fig:small-agg-examples} and
\ref{fig:large-agg-examples}. \citet{Westbrook-2004a,Westbrook-2004b}
showed that the mean aspect ratio of aggregate snowflakes approaches
an asymptotic value of 0.65 for large $N$ in their data. This
asymptotic value is independent of the monomer
shape. Figure~\ref{fig:aspect-ratio} confirms their finding, but with
a slightly larger asymptotic value of 0.74 in our data, suggesting
more compact growth in our model at large $N$. This discrepancy may be
due to differences in the aggregation model, such as grid resolution
and, in particular, the concept of a penetration depth in our model,
which may allow for more compact growth compared to that
in \citet{Westbrook-2004a,Westbrook-2004b}. The main difference,
however, is that Westbrook et al. define the aspect ratio as an
average over random projections including the vertical axis resulting
in a smaller absolute value.

As expected from the aspect ratios of their respective monomers,
aggregates of plates begin with a mean aspect ratio close to 0.9 (plates have a horizontal aspect ratio of one), with the aspect
ratio decreasing monotonically with $N$ toward 0.73
(Fig.~\ref{fig:aspect-ratio}a). In contrast, aggregates of needles
start around 0.5, with $\phi$ increasing monotonically with $N$.  For
aggregates of needles the relation for the mean aspect ratio as a
function of $N$ can be parameterized with:
\begin{equation} \label{eq:phi0}
  \phi_0(N) = 0.742 - 0.227  \, \left ( \frac{2}{N} \right )^{0.95}
\end{equation}
and for plates aggregates with
\begin{equation}
  \phi_1(N) = 0.736 + 0.141  \, \left ( \frac{2}{N} \right )^{0.65}
\end{equation}
The data of the 50/50 mixture aggregates of needles and plates follow the non-monotonic relation
\begin{equation}
  \phi\halfP(N) =
  \begin{cases}
    0.751 - 0.0456 \left (\frac{2}{N} \right)^{1.5},     & \text{for } N < 32, \\
    0.733 + 0.0186 \left ( \frac{32}{N} \right )^{0.3},  & \text{otherwise.}    
  \end{cases}
\end{equation}
For other values of $\MR$, linear interpolation between these three relations provides a sufficiently good approximation.

For aggregates of dendrites and the mixture of needles and dendrites
the aspect ratio behaves very similar to the plate case. The only
difference is that at low $N$ aggregates of dendrites have a slightly
larger standard deviation than aggregates of plates. The corresponding parameterizations are:
\begin{equation}
  \phi_2(N) = 0.732 + 0.172  \, \left ( \frac{2}{N} \right )^{0.6}
\end{equation}
and
\begin{equation}  \label{eq:phi_mix2}
  \phi\halfD(N) =
  \begin{cases}
    0.756 - 0.0416 \left (\frac{2}{N} \right)^{2},       & \text{for } N < 16, \\
    0.727 + 0.0271 \left ( \frac{16}{N} \right )^{0.3},  & \text{otherwise.}    
  \end{cases}
\end{equation}

\def\mywidth{7.8cm}
\begin{figure}[t]
  \centering
  \begin{minipage}{\mywidth}
    a) aggregates of needles and plates \\
    \includegraphics[width=\textwidth,viewport= 50 150 490 600,clip=]
                    {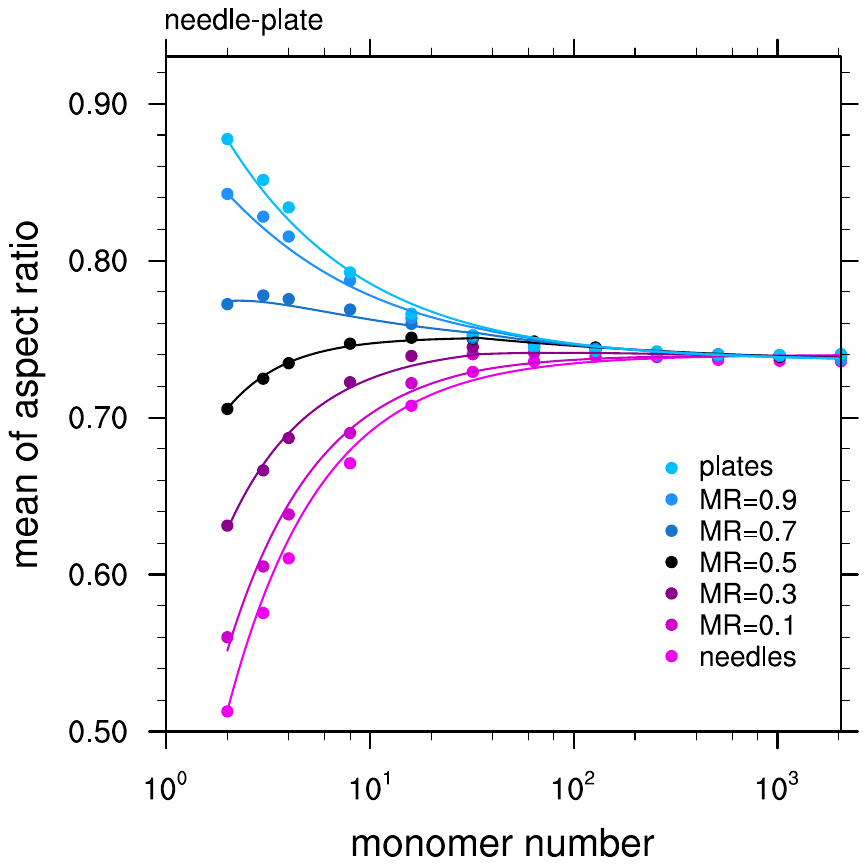}
  \end{minipage}
  \hfill
  \begin{minipage}{\mywidth}
    b) aggregates of needles and dendrites \\
    \includegraphics[width=\textwidth,viewport= 50 150 490 600,clip=]
                    {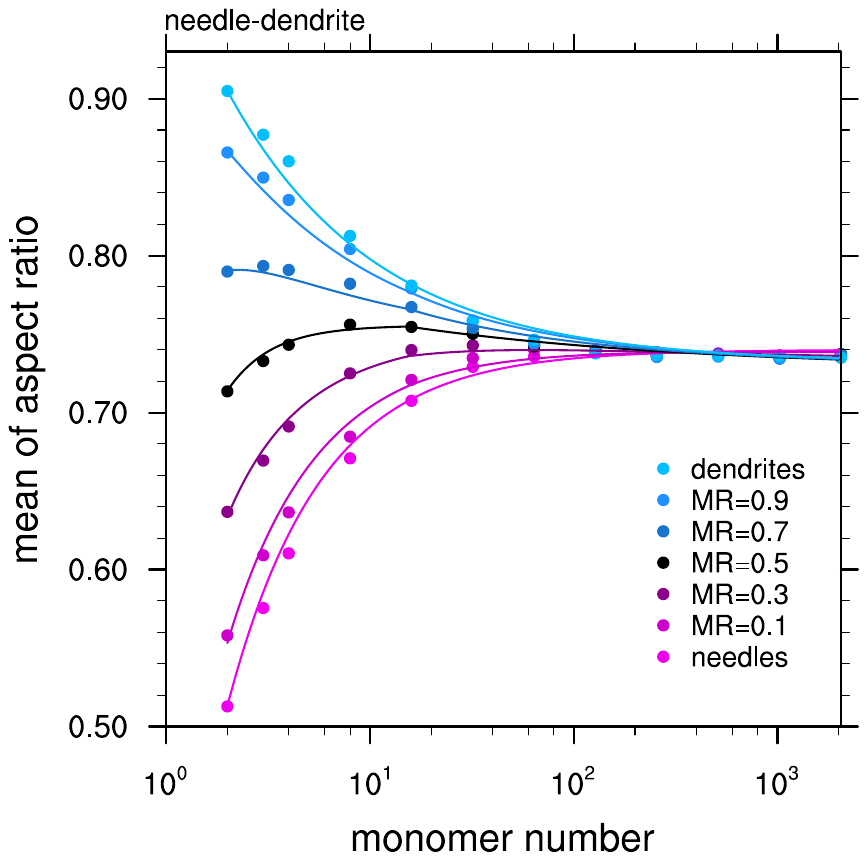}
  \end{minipage}  
  \caption{Mean horizontal aspect ratio $\phi_h$ of aggregates of
    needles and plates (left) and aggregates of needles and dendrites
    (right) as a function of monomer number $N$. Aggregates of
    mixtures are shown for different monomer ratios $\MR$. Note that
    the data for pure needle aggregates is the same in both plots.
    The solid lines are parameterizations using Eqs.~\eqref{eq:phi0}–\eqref{eq:phi_mix2},
    with intermediate values linearly interpolated between
    $\MR=0$ (needles), $\MR=1$ (plates or dendrites), and $\MR=0.5$.
    \label{fig:aspect-ratio}}
\end{figure}

\subsection{Cross-sectional area}

The cross-sectional area is a key property of hydrometeors, as it
influences aerodynamic drag, terminal fall velocity, and the collision
kernel \citep{Boehm-1992a,Boehm-1992b,Boehm-1992c}.  Following Böhm's
formulation, we define the ellipsoidal area ratio $q$ as the ratio of
the snowflake’s projected area to that of its circumscribing ellipse.

Figure~\ref{fig:qarea} shows that $q$ decreases with increasing
monomer number $N$.  As aggregates grow, they become more fractal and
less dense, reducing their area ratios.
In contrast to the aspect
ratio $\phi$, the asymptotic value of $q$ depends on the monomer
shape, as the internal structure of the aggregate contributes to the
projected area.

For example, aggregates of needles approach a asymptotic value of
approximately 0.33, while aggregates of plates approach level off
around 0.41 (Fig.~\ref{fig:qarea}a).  Because $q$ accounts for
internal structure, not just the outer envelope like $\phi$, monomer
geometry remains influential at all sizes. In other words, each small
crystal within the aggregate contributes to the total cross-sectional
area, preserving a dependence on monomer shape.

The parameterizations of $q_\MR$ for aggregates of
needles, aggregates of plates and aggregates of the 50/50 mixture are:
\begin{equation} \label{eq:q0}
  q_0(N) =  0.327 + 0.386 \left ( \frac{2}{N} \right )^{0.75}
\end{equation}
and
\begin{equation}
  q_1(N) =
  \begin{cases}
    0.900 - 0.305 \left ( \frac{2}{N}\right )^{1.5},     & \text{for } N < 32, \\
    0.378 + 0.217 \left ( \frac{32}{N} \right )^{0.45},  & \text{otherwise.}
  \end{cases}
\end{equation}
and
\begin{equation}
  q\halfP(N) =  0.346 + 0.331 \left ( \frac{4}{N} \right )^{0.35}.
\end{equation}
Again, linear interpolation applies for other monomer ratios $\MR$.

Figure~\ref{fig:qarea}b confirms that monomer shape influences the
area ratio $q$, as seen in aggregates of dendrites and their mixtures.
Aggregates of dendrites have an even lower area ratio $q$
than aggregates of needles, at least for $N<10$. For intermediate $N$
aggregates of dendrites show a slightly higher $q$ until they converge
to a very similar value for large $N$. The parameterization for the
area ratio $q$ of aggregates of dendrites is
\begin{equation}
  q_2(N) =
  \begin{cases}
    0.404 - 0.0252 \left ( \frac{2}{N}\right )^{2},     & \text{for } N < 16, \\
    0.178 + 0.2660 \left ( \frac{16}{N} \right )^{0.1},  & \text{otherwise.}
  \end{cases}
\end{equation}
and for the mixture aggregates of needles and dendrites
\begin{equation}  \label{eq:q_mix2}
  q\halfD(N) =  0.315 + 0.173 \left ( \frac{3}{N} \right )^{0.45}.
\end{equation}

\def\mywidth{7.8cm}
\begin{figure}[t]
  \centering
  \begin{minipage}{\mywidth}
    a) aggregates of needles and plates \\
    \includegraphics[width=\textwidth,viewport= 50 150 490 600,clip=]
                    {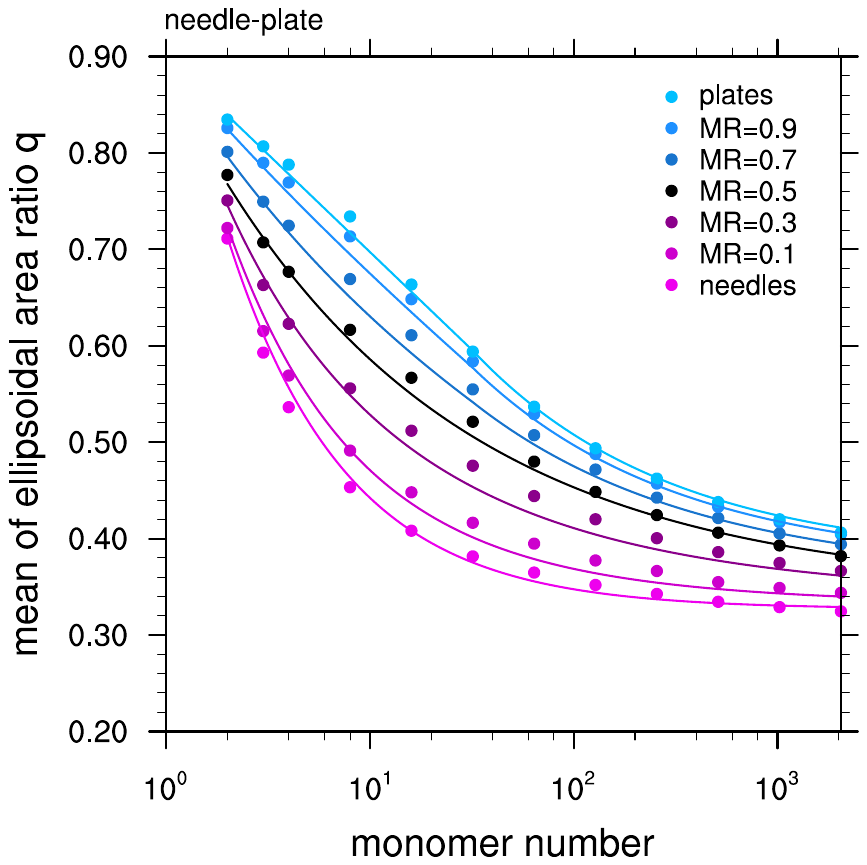}
  \end{minipage}
  \hfill
  \begin{minipage}{\mywidth}
    b) aggregates of needles and dendrites \\
    \includegraphics[width=\textwidth,viewport= 50 150 490 600,clip=]
                    {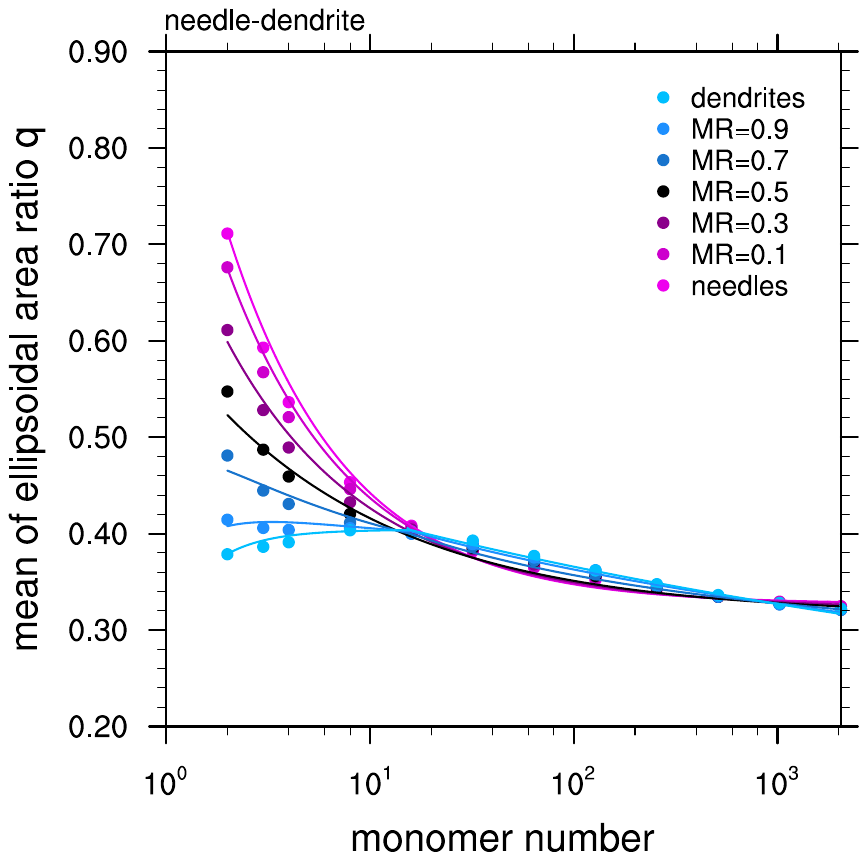}
  \end{minipage}  
  \caption{As Figure \ref{fig:aspect-ratio} but for the mean
    ellipsoidal area ratio $q$. The solid lines are the
    parameterizations using Eqs.~\eqref{eq:q0}-\eqref{eq:q_mix2}.
    \label{fig:qarea}}
\end{figure}

These parameterizations of snowflake geometry in terms of $\Dmax$,
$\phi$, and $q$ provide a physically grounded framework for both improving
microphysical representations in Lagrangian particle models and
enabling more accurate retrievals from remote sensing
observations. The required input variables are the mass $m$, the
monomer number $N$, the monomer ratio $\MR$, and the mass-size
relations of plate, needle, and dendrite monomers.

\subsection{Joint probability functions}

\begin{figure}[t]
  \centering

  \begin{minipage}{0.32\linewidth}
    {\small a) agg. of needles with $N = 64$} \\[3mm]
    \includegraphics[width=\linewidth,
      viewport=70 150 510 590,clip]
      {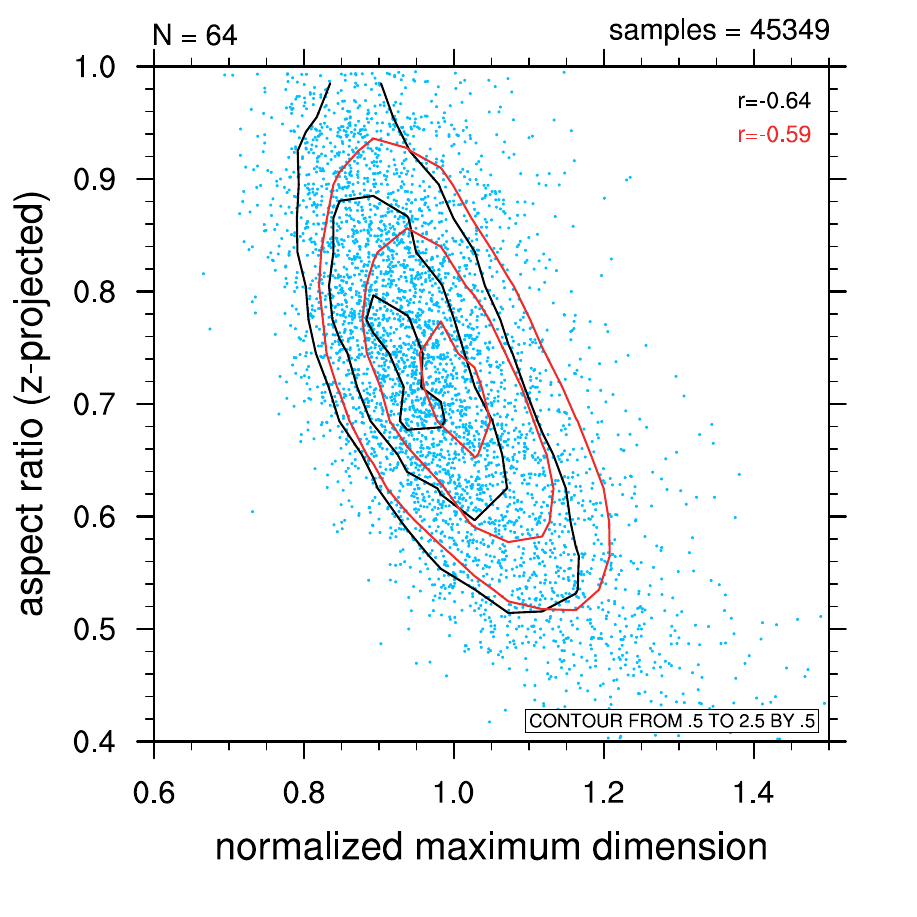}
  \end{minipage}%
  \hfill%
  \begin{minipage}{0.32\linewidth}
    {\small b) agg. of plates with $N = 64$} \\[3mm]
    \includegraphics[width=\linewidth,
      viewport=70 150 510 590,clip]
      {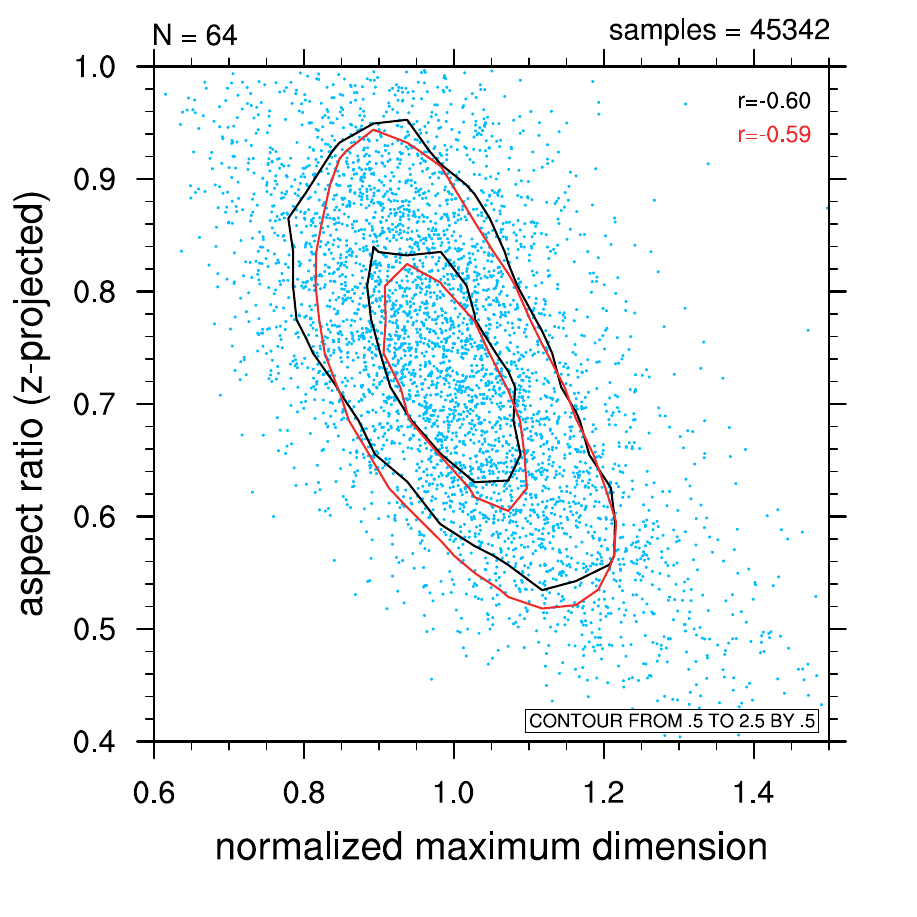}
  \end{minipage}%
  \hfill%
  \begin{minipage}{0.32\linewidth}
    {\small c) agg. of dendrites with $N = 64$} \\[3mm]
    \includegraphics[width=\linewidth,
      viewport=70 150 510 590,clip]
      {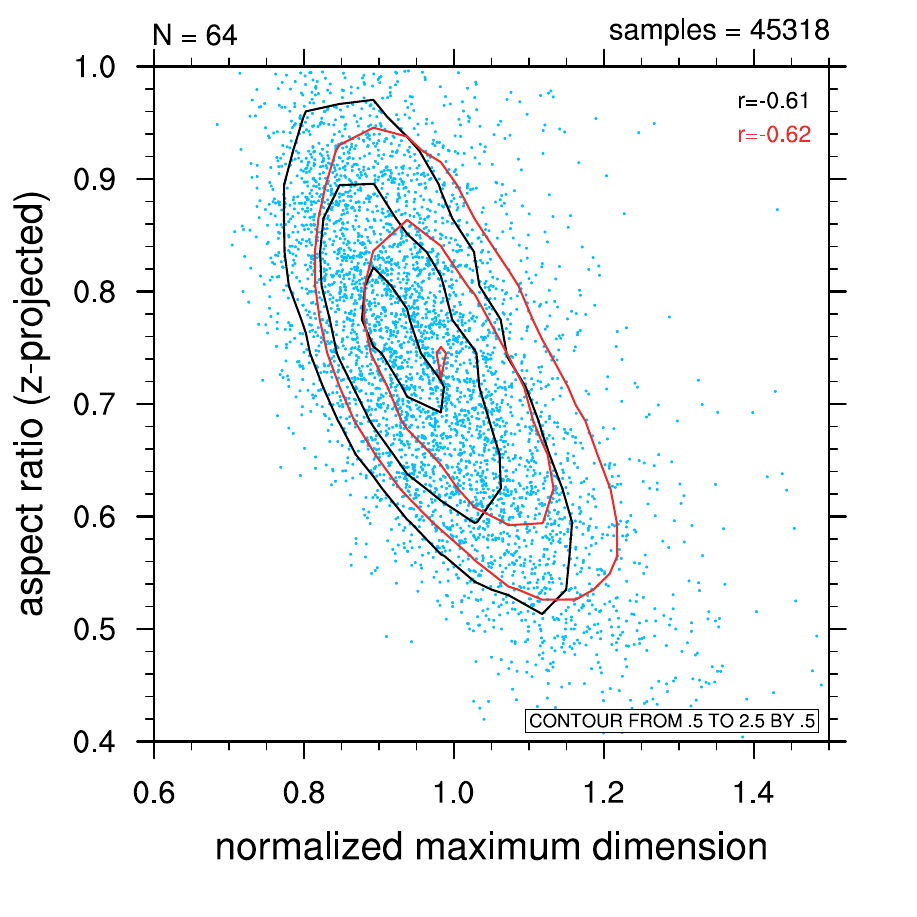}
  \end{minipage}

  \vspace{4mm}

  \begin{minipage}{0.32\linewidth}
    {\small d) agg. of needles with $N = 1024$} \\[3mm]
    \includegraphics[width=\linewidth,
      viewport=70 150 510 590,clip]
      {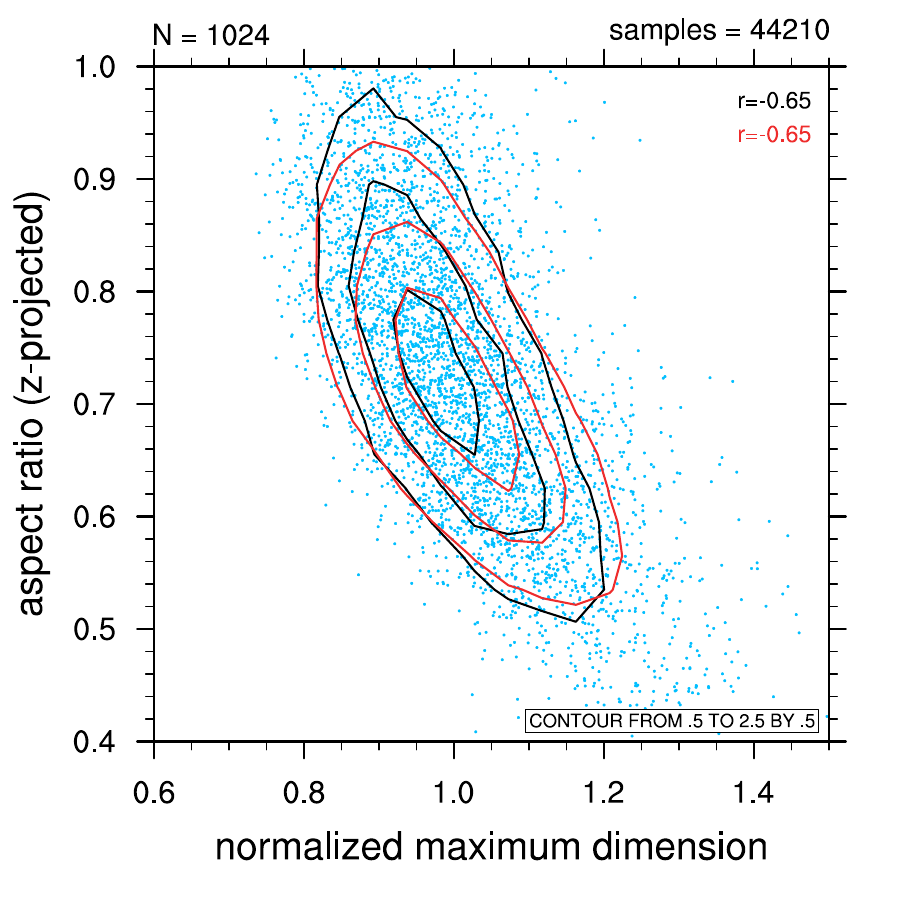}
  \end{minipage}%
  \hfill%
  \begin{minipage}{0.32\linewidth}
    {\small e) agg. of plates with $N = 1024$} \\[3mm]
    \includegraphics[width=\linewidth,
      viewport=70 150 510 590,clip]
      {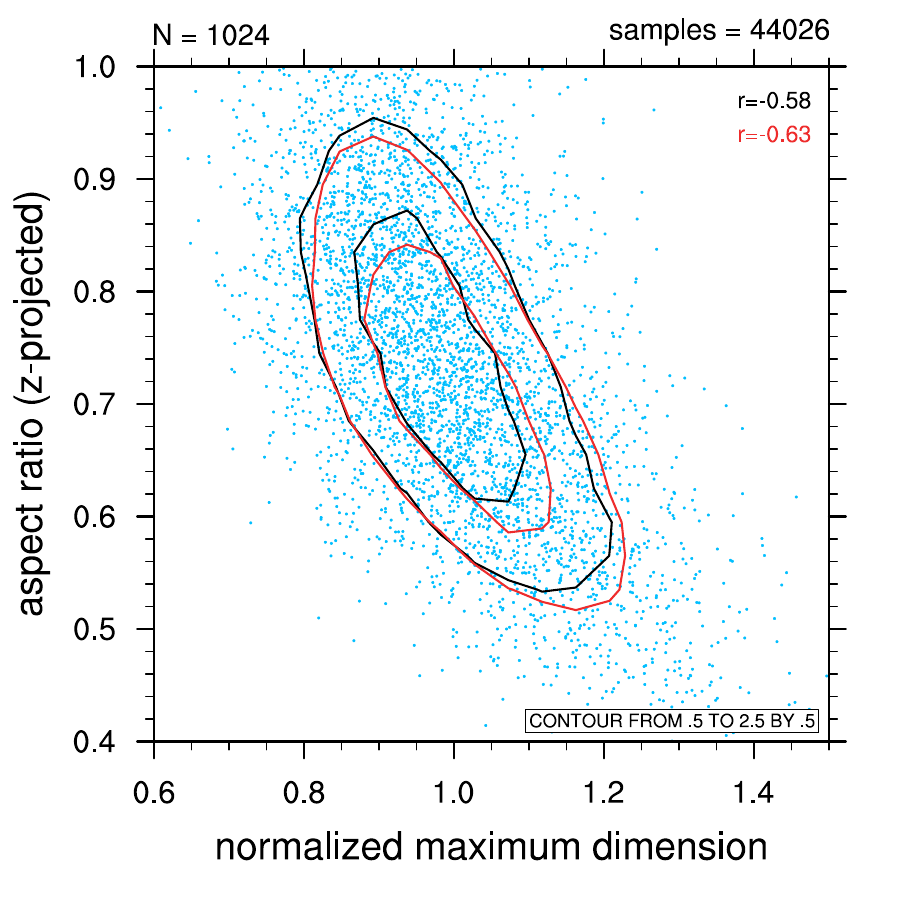}
  \end{minipage}%
  \hfill%
  \begin{minipage}{0.32\linewidth}
    {\small f) agg. of dendrites with $N = 1024$} \\[3mm]
    \includegraphics[width=\linewidth,
      viewport=70 150 510 590,clip]
      {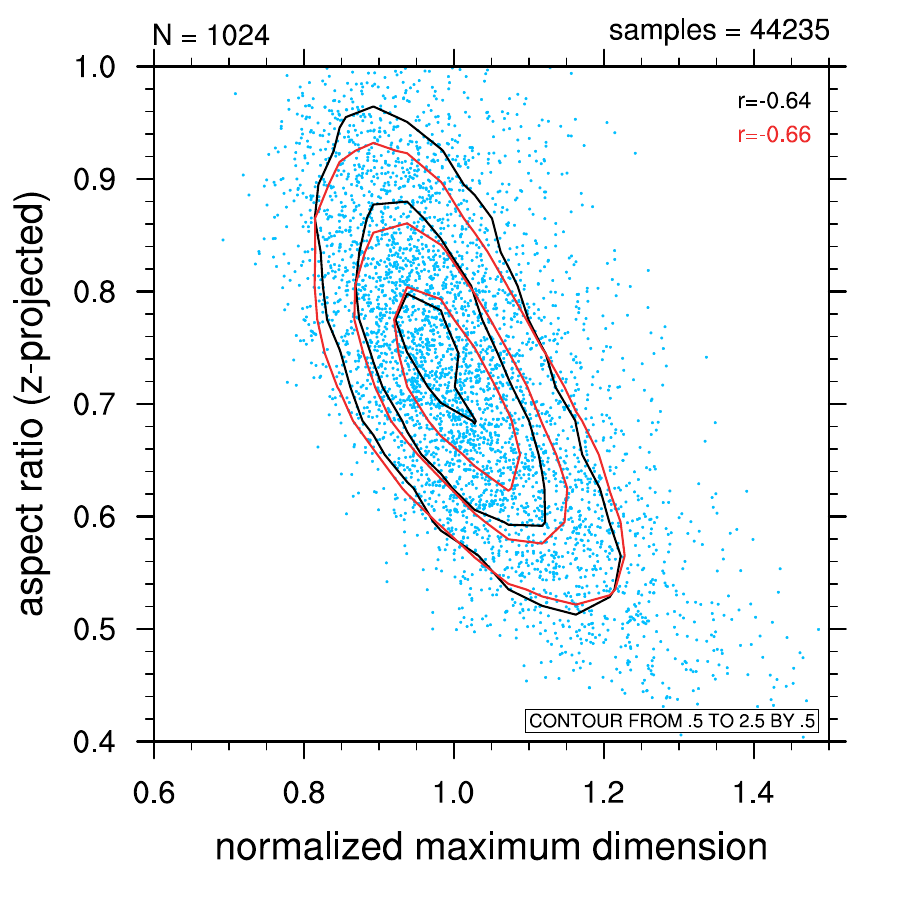}
  \end{minipage}

  \caption{Joint PDF of the normalized maximum dimension $\Dnorm$ and
    the aspect ratio $\phi$ for different monomer habits and monomer
    numbers. The blue dots show thinned-out samples of the dataset,
    black isolines correspond to the data. The solid red lines are the
    parameterizations using Eqs.~\eqref{eq:phi_pdf}-\eqref{eq:sig_psi}. The correlation coefficient
    is given in the upper right corner, in black for the data and in
    red for the parameterization. 
    \label{fig:jointpdf}}
\end{figure}

For a stochastic model of snowflake geometry, it is not sufficient to
describe individual properties such as $\Dmax$, $\phi$, and $q$ in
isolation; we must also parameterize the correlations between these
variables to generate self-consistent and realistic aggregate
snowflakes.  In particular, a strong correlation exists between the
normalized maximum dimension $\Dnorm$ and the aspect ratio $\phi$, as
illustrated by the snowflake examples in
Figs.~\ref{fig:small-agg-examples} and \ref{fig:large-agg-examples}
and confirmed in Fig.\ref{fig:jointpdf}, which shows their joint
probability density function (PDF).

In the tails of the distributions, values of $\phi > 1$ appear due to
the use of the principal axis transform (based on the inertia tensor)
and the definition of $\phi$ via geometric vertical projections.  This
does not affect the results, as values with $\phi > 1$ are later
mapped to $1/\phi$, effectively ensuring that $\phi \leq 1$ and
reversing the identity of the two horizontal axes.

The joint PDFs of aggregates of needles exhibits an elongated shape, with
correlation coefficients of $r = -0.64$ for $N = 64$ and $r = -0.65$
for $N = 1024$. Aggregates of plates have slightly lower correlation
coefficients with of $r = -0.60$ for $N = 64$ and $r = -0.58$ for $N =
1024$. Aggregates of dendrites fall in between with $r = -0.61$ for $N
= 64$ and $r = -0.64$ for $N = 1024$.

The joint distribution of $\phi$ and $\Dnorm$ can be approximated by
the product of a lognormal distribution for $P(\Dnorm)$ and a
lognormal conditional distribution for $P(\phi \mid \Dnorm)$:
\begin{equation}
  P(\phi, \Dnorm) = P(\Dnorm) \, P(\phi \mid \Dnorm).
\end{equation}
Rather than specifying closed-form expressions for the probability
density functions, we define stochastic generation rules suitable for
direct implementation in a Monte Carlo framework:
\begin{align} \label{eq:phi_pdf}
  &P(\Dnorm) = \exp \left[ - \frac{1}{2} \sigmaD^2 + \mathcal{N}(0, \sigmaD^2) \right], \\
  &P(\phi \mid \Dnorm) = \exp \left[ \log\left( \psi(\Dnorm) \bar{\phi}(N) \right ) - \frac{1}{2} \sigma_\phi^2 + \mathcal{N}(0, \sigma_\phi^2) \right],
\end{align}
where $\mathcal{N}(0,\sigma^2)$ denotes a normally distributed random
variable with zero mean and variance $\sigma^2$. This approach allows
efficient sampling of aggregate geometries with physically consistent
correlations.  The function $\psi(\Dnorm)$ introduces a size-dependent
skewness to the conditional distribution:
\begin{equation} \label{eq:psi_pdf}
  \psi(\Dnorm) =
  \begin{cases}
    0.55 + 0.45 \, \Dnorm^{-3}, & \Dnorm > 1 \\
    1.45 - 0.45 \, \Dnorm^{2},  & \text{otherwise.}
  \end{cases}
\end{equation}
By making the location parameter of the lognormal distribution - the
mean in log-space - a function of $\Dnorm$, the conditional
distribution of $\phi$ (and similarly $q$ below) becomes skewed in a
physically meaningful way. For instance, larger aggregates may tend to
appear more elongated or irregular in the horizontal plane, a behavior
that can be captured by an appropriate choice of the function
$\psi(\Dnorm)$.  When combined with the parameterization for
$\sigmaD$, this results in the joint PDF shown by the red isolines in
Fig.~\ref{fig:jointpdf}, yielding correlation coefficients of $r =
-0.60$ for $N = 64$ and $r = -0.65$ for $N = 1024$ for aggregates of
needles; $r = -0.60$ for $N = 64$ and $r = -0.64$ for $N = 1024$ for
aggregates of plates; and $r = -0.63$ for $N = 64$ and $r = -0.65$ for
$N = 1024$ for aggregates of dendrites. Given the uncertainties of the
aggregation model, the accuracy of these parameterizations should be
sufficient. Interestingly, the functional form of $\psi$ is
independent from the monomer habit and the same relation can be used
for all aggregates. The standard deviation is given by
\begin{equation} \label{eq:sig_psi}
  \sigma_\phi = 0.12 + \frac{0.2}{\sqrt{N}}
\end{equation}
independent from monomer habit.

For the joint probability of $\Dnorm$ and the ellipsoidal area ratio
$q$ we use the same ansatz with a lognormal for the conditional
distribution
\begin{equation} \label{eq:q_pdf}
  P(q \mid \Dnorm) = \exp \left[ \, \log \left( \, \chi_\MR(\Dnorm) \, \bar{q}(N) \, \right) - \frac{1}{2} \sigma_q^2  + \mathcal{N}(0, \sigma_q^2) \, \right],
\end{equation}
with $\sigma_{q,0} = 2.07$, $\sigma_{q,1} = 0.127$, $\sigma_{q,2} = 0.180$, $\sigma_{q,\half}^\plate = 0.144$
for the 50/50 mixture of needles and plates, and
$\sigma_{q,\half}^\plate = 0.168$ for the 50/50 mixture of needles and
dendrites.

For aggregates of needles $\chi_0(\Dnorm)$ is given by:
\begin{equation}  \label{eq:chi0_pdf}
  \chi_0(\Dnorm) =
  \begin{cases}
    0.70 + 0.20 \, \Dnorm^{-2} \, & \Dnorm > 1 \\
    1.25 - 0.35 \, \Dnorm^{3} \, & \text{otherwise.}
  \end{cases}
\end{equation}
and for all other cases, plates, dendrites, and both types for mixture, the following approximation applies:
\begin{equation}  \label{eq:chi1_pdf}
  \chi_1(\Dnorm) = \chi_2(\Dnorm) = \chi_\half(\Dnorm) =
  \begin{cases}
    0.75 + 0.20 \, \Dnorm^{-2} \, & \Dnorm > 1 \\
    1.40 - 0.45 \, \Dnorm^{3} \, & \text{otherwise.}
  \end{cases}
\end{equation}
The joint probability density functions (PDFs) of $\Dnorm$ and $q$ for
pure aggregates of needles, plates and dendrites are shown in
Fig.~\ref{fig:jointpdf_qarea}.  For aggregates of plates, the
correlation between $q$ and $\Dnorm$ is moderately strong. The
aggregate dataset yields correlation coefficients of $r = -0.49$ for
$N = 64$ and $r = -0.55$ for $N = 1024$
(Figs.~\ref{fig:jointpdf_qarea}b,e). The parameterization performs
reasonable, but with slightly higher correlation of $r = -0.57$ for
both $N$. Aggregates of needles exhibit a weaker relationship. In the
data, the correlation drops to $r = -0.46$ for $N = 64$ and $r =
-0.29$ for $N = 1024$ (Figs.~\ref{fig:jointpdf_qarea}a,d). The
parameterization captures this trend, yielding $r = -0.35$ for both
values of $N$. Especially for the case of aggregates of needles with
$N=64$ (Fig.~\ref{fig:jointpdf_qarea}c) the strong skewness in $q$
observed for $\Dnorm < 1$ is not fully captured by the simple ansatz
for the joint PDF. Consequently, the parameterization underestimates
the probability of needle aggregates with a high $q$ in the tail of
the distribution. Aggregates of dendrites are somewhat in between and
show a significant but weaker nonlinear tail.  Aggregates of mixtures
tend to resemble their more oblate component in terms of $q$;
corresponding figures are omitted for brevity.

While our simple joint PDF model captures general trends and the
essential behavior, it is not able to describe the full complexity of
aggregate snowflakes in a quantitative form. For small $N$, and
especially for $N<10$, the variability in $q$ as reflected in the
joint PDF is higher and more strongly influenced by individual monomer
geometry. This complexity for small $N$ may warrant a more detailed
treatment in future work.



\def\mywidth{5.4cm}
\begin{figure}[t]
  \centering
  \begin{minipage}{\mywidth}
    {\small a) agg. of needles with $N = 64$} \\[3mm]
    \includegraphics[width=\textwidth,viewport= 70 150 510 590,clip=]
                    {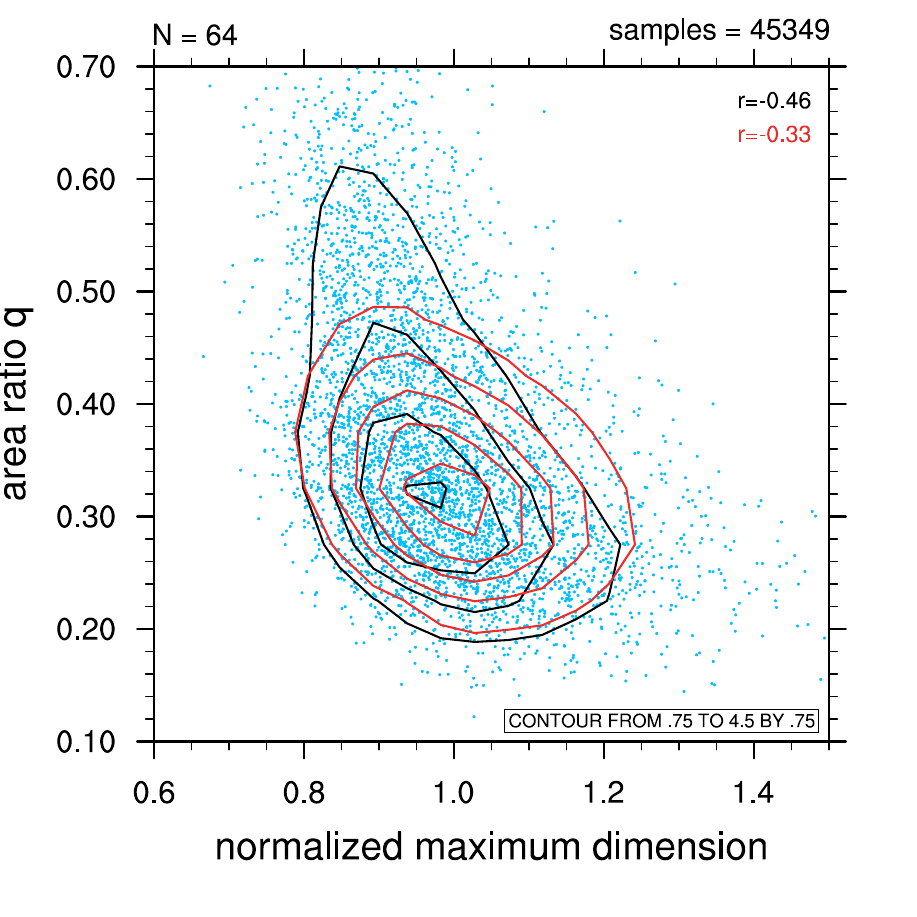}
  \end{minipage}
  \begin{minipage}{\mywidth}
    {\small b) agg. of plates with $N = 64$} \\[3mm]
    \includegraphics[width=\textwidth,viewport= 70 150 510 590,clip=]
                    {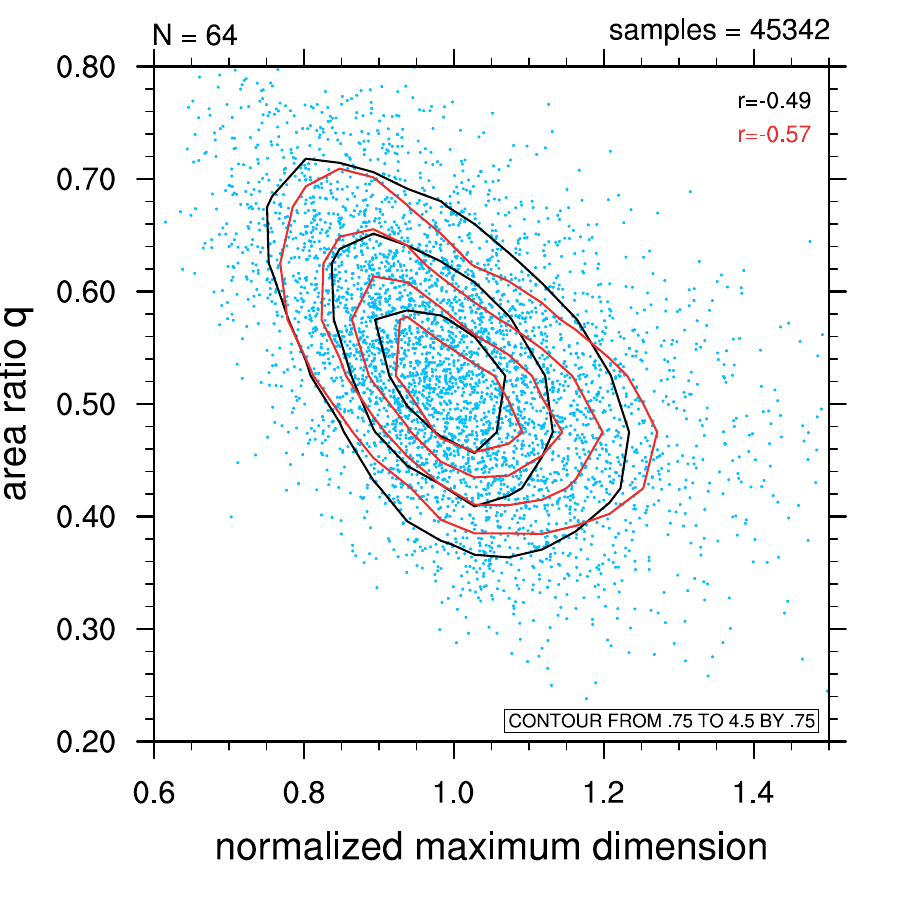}
  \end{minipage}
  \begin{minipage}{\mywidth}
    {\small c) agg. of dendrites with $N = 64$} \\[3mm]
    \includegraphics[width=\textwidth,viewport= 70 150 510 590,clip=]
                    {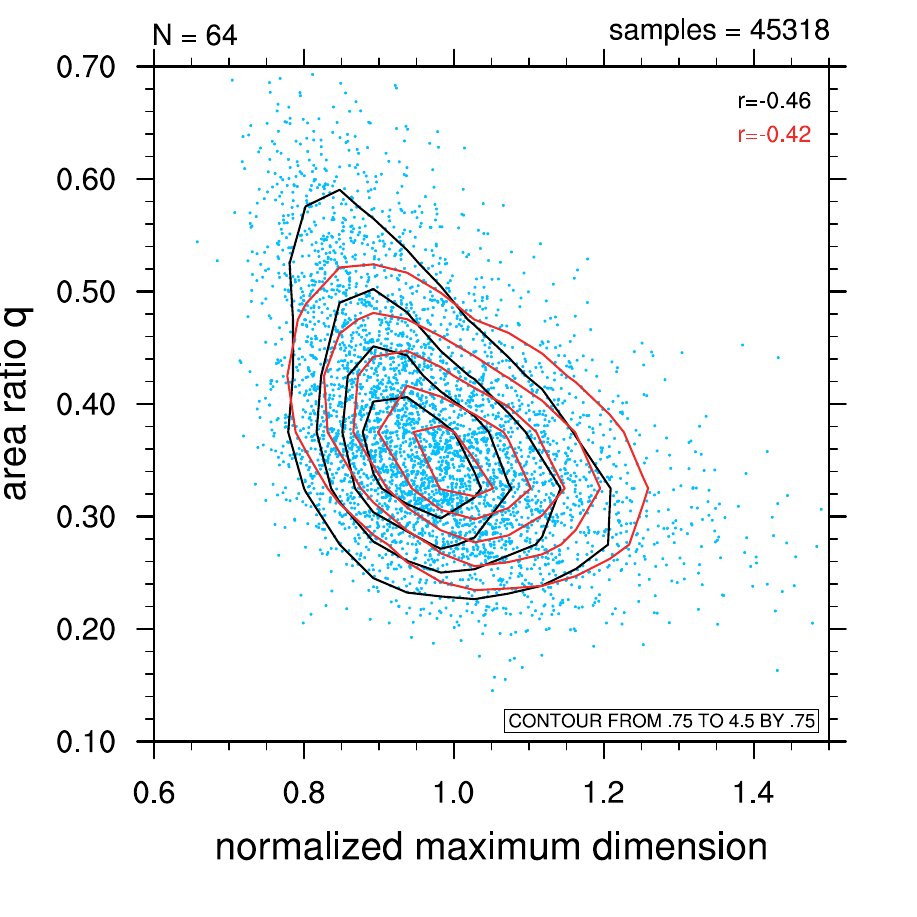}
  \end{minipage}
  
  \vspace{3mm}
  \begin{minipage}{\mywidth}
    {\small d) agg. of needles with $N = 1024$} \\[3mm]
    \includegraphics[width=\textwidth,viewport= 70 150 510 590,clip=]
                    {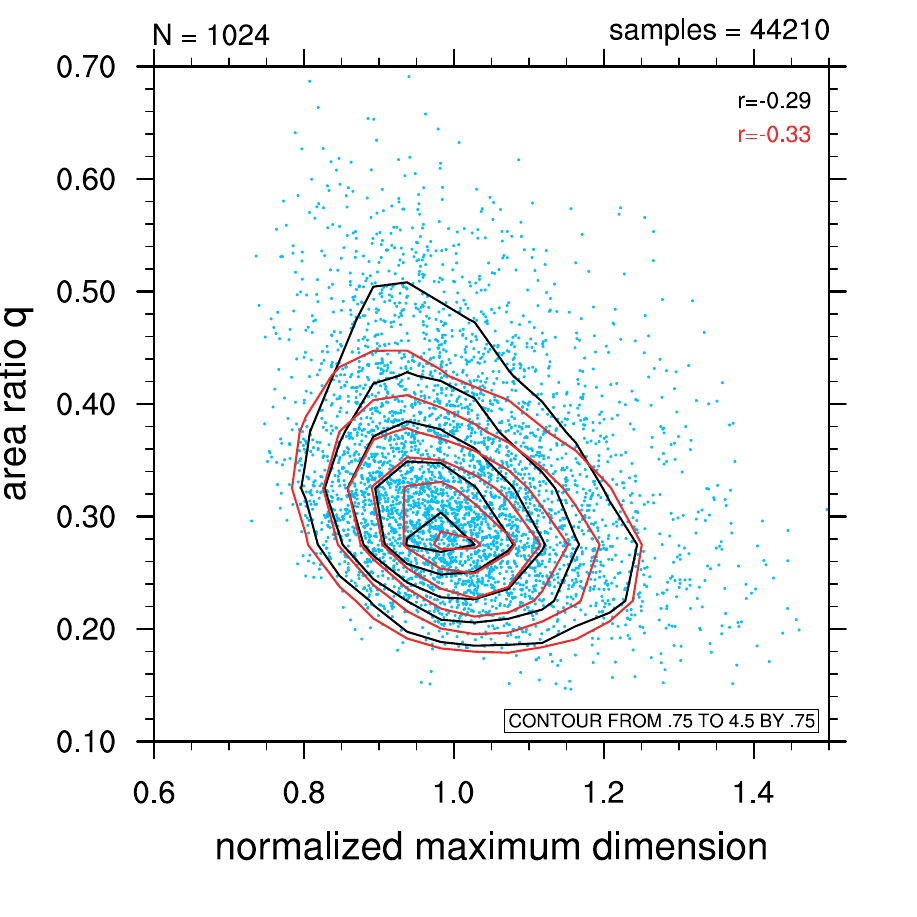}
  \end{minipage}  
  \begin{minipage}{\mywidth}
    {\small e) agg. of plates with $N = 1024$} \\[3mm]
    \includegraphics[width=\textwidth,viewport= 70 150 510 590,clip=]
                    {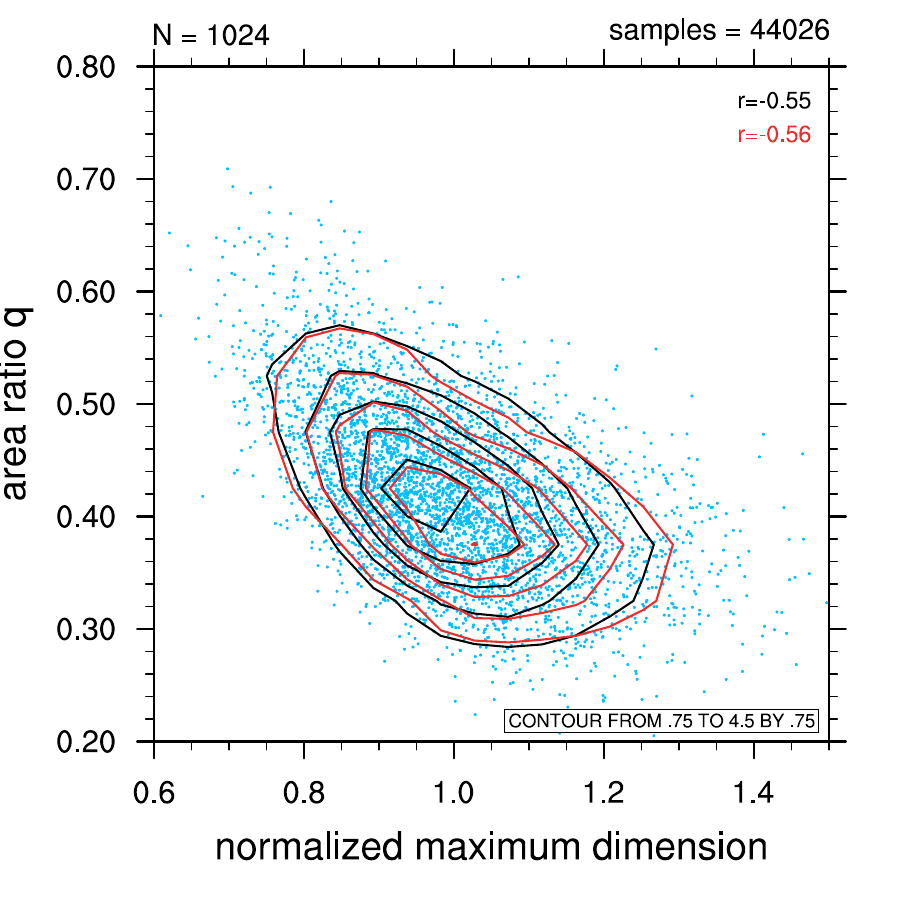}
  \end{minipage}
  \begin{minipage}{\mywidth}
    {\small f) agg. of dendrites with $N = 1024$} \\[3mm]
    \includegraphics[width=\textwidth,viewport= 70 150 510 590,clip=]
                    {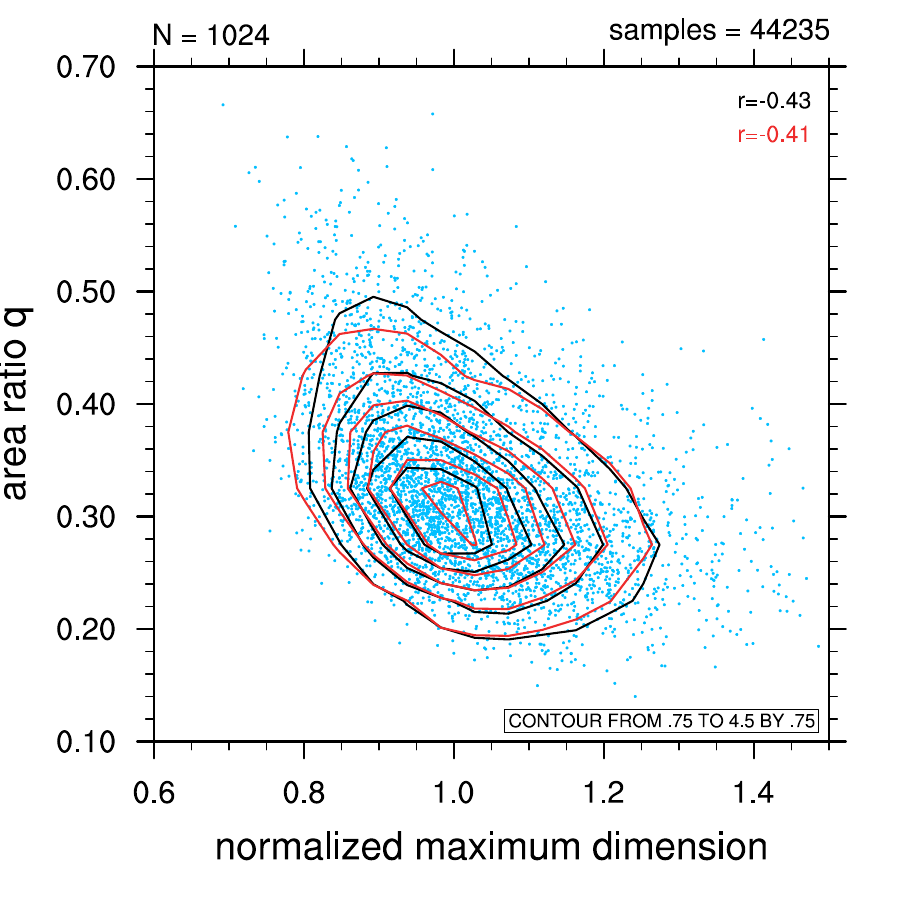}
  \end{minipage}
  
  \caption{As Fig.~\ref{fig:jointpdf_qarea}, but for the joint
    correlation of the normalized maximum dimension $\Dmax$ and the
    ellipsoidal area ratio $q$. The solid red lines are the
    parameterizations using Eqs.~\eqref{eq:q_pdf}-\eqref{eq:chi1_pdf}.
    \label{fig:jointpdf_qarea}}
\end{figure}

\section{Application in McSnow}\label{sec:apl_mcsnow}

\subsection{Implementation}

The Monte Carlo Lagrangian Particle Model \McSnow\ predicts eight key
properties of hydrometeors. These properties include the ice mass $m$,
the monomer number $N$ of aggregates, the rime mass $\mrime$, the rime volume
$\vrime$, the liquid mass $\mliq$, and the frozen mass $\mfrz$. In
addition, when predicting the habit of primary crystals (or monomers),
the model introduces the prognostic variables of the monomer aspect
ratio $\phi_m$ and the ice volume $\vice$ \citep{Welss-2024}, with the
latter being essential for characterizing branching and hollowing of
crystals. Furthermore, the multiplicity $\xi$ is used to describe the
number of real particles represented by a super-particle
\citep{Shima-2009}.

While the maximum dimension and other geometric properties, such as
the ellipsoidal area ratio $q$, are not prognostic variables in
\McSnow\ at present, they can be diagnosed from the eight prognostic
variables, combined with assumptions about particle shape.  For
instance, \citet{Seifert-2019} discuss geometry assumptions for
partially rimed snowflakes and graupel. In this work, we extend the
framework to include stochastic aggregates with habit information,
leveraging the joint PDFs and geometric relationships discussed in the
previous section. The maximum dimension $\Dmax$ for aggregate
snowflakes will be treated as a stochastic pseudo-prognostic variable,
as explained in the following.

The stochastic aggregate snowflake framework described earlier is now
integrated into \McSnow. In addition to the monomer number $N$, we
introduce a monomer count for prolate crystals, $N_p$, which simply
counts the monomers in an aggregate with $\phi_m > 1$. Using this, the
monomer ratio $\mr$ for an individual aggregate is calculated as:
\begin{equation} \label{eq:mcsnow-mr}
  \mr = \frac{N_p}{N}
\end{equation}
At first glance, using the monomer ratio $\mr$ of an individual
aggregate may appear inconsistent with Section 3, where $\MR$ is
derived from the size distribution of primary crystals. However, in
the context of a super-particle, $\mr$ naturally represents an
ensemble average over the $\xi$ real particles it stands for, thereby
resolving the apparent inconsistency.

In addition, the number of dendrite monomers in each aggregate is
tracked using the counter $N_d$. Dendrite monomers are defined as
oblate monomers with an ellipsoidal area fraction $q$ smaller than 0.5
at the time of collision. Given $N$, $N_p$, and $N_d$, the number of plate monomers in an aggregate is $N_o = N -
N_p - N_d$. If $N_d > N_o$, the aggregate is treated as a mixture of
needles and dendrites; otherwise, as a mixture of needles and
plates. This simplification avoids complicated multidimensional
interpolation in a three-component mixture and neglects the existence
of plate–dendrite combinations. This approach can be generalized if
observations show that such mixtures are more relevant than currently
assumed.

After a successful collision event between ice crystals and/or
aggregate snowflakes, the geometry of the resulting new aggregate has
to be determined. Given the mass $m$, the monomer number $N$, and the
monomer ratio $\mr$ of the new aggregate, the mean monomer size
$\Dmono$ is computed using Eq.~\eqref{eq:Dmono}. For aggregates of
mixtures the geometry coefficients of the monomers are interpolated
with:
\begin{align}
  a_\mr &= \exp ( \, \mr \,\log(a_1) + (1-\mr) \, \log(a_0) \, )\\
  b_\mr &= \mr \, b_1 + (1-\mr) \, b_0. 
\end{align}
The normalized maximum dimension for the new aggregate is then sampled
from the following log-normal distribution:
\begin{equation}
  \Dnorm \sim  \exp \left[ - \frac{\sigmaD^2}{2} + \mathcal{N}(0, \sigmaD^2) \right],
\end{equation}
where $\sim$ denotes that a random variate is drawn from the
distribution specified on the right-hand side. The maximum dimension
$\Dmax$ is then computed by inverting Eq.~\eqref{eq:Dnorm}, yielding:
\begin{equation}
  \Dmax = \Dnorm \, \Dmono \left ( \frac{N}{N_0} \right )^\eta + \Dmin.
\end{equation}
Remember that $\Dnorm$ is dimensionless, which ensures the dimensional
consistency of this equation. Once $\Dnorm$ is sampled, corresponding
random variates for the aspect ratio $\phi$ and the area ratio $q$ are
drawn from the following distributions:
\begin{equation}
\phi \sim \exp \left[ \log\left( \psi(\Dnorm) \, \bar{\phi}(N) \right)  - \frac{1}{2} \sigma_\phi^2 + \mathcal{N}(0, \sigma_\phi^2) \right],
\end{equation}
and
\begin{equation}
q \sim \exp \left[ \log \left( \chi_\mr(\Dnorm) \, \bar{q}(N) \right) - \frac{1}{2} \sigma_q^2 + \mathcal{N}(0, \sigma_q^2) \right].
\end{equation}
With the geometric properties $\Dmax$, $\phi$, and $q$ available, the
terminal fall velocity and collision efficiency of the newly formed
aggregate can be calculated using Böhm's parameterizations
\citep{Boehm-1992a, Boehm-1992b, Boehm-1992c}, as described and
concisely summarized in Appendix A of \citet{Welss-2024}. These
parameters, together with the underlying geometric assumptions,
provide the basis for computing the collision kernel in subsequent
aggregation or riming events, ensuring consistent microphysical
treatment. The random variates for $\Dmax$, $\phi$, and $q$ assigned
to each Monte Carlo aggregate are stored and continuously updated to
reflect changes caused by microphysical processes other than collision. For
depositional growth, for example, the update of $\Dmax$ is calculated
as
\begin{equation}
  \Dmax^{\text{new}} = \Dmax^{\text{old}} \left( \frac{m + \Delta m_\text{depo}}{m} \right)^\zeta
\end{equation}
where $\Delta m_\text{depo}$ is the mass gain from deposition in the current time step, and $\zeta$ is defined as
\begin{equation}
  \zeta = \frac{1}{N} \left( \frac{1}{b_h} + \frac{N-1}{b_\text{agg}} \right).
\end{equation}
and $b_h$ is again the interpolated monomer geometry exponent and
$b_\text{agg}=2.1$.
This ansatz accounts for the geometric influence of monomers at low
$N$ via the exponent $\beta_i$, while ensuring that $\zeta$
asymptotically approaches $1/b_\text{agg}$ for large aggregates.

During depositional growth, $\phi$ and $q$ are held constant. While
this lacks direct physical validation, it provides a practical and
robust approximation given the scarcity of detailed observational or
simulation data. Importantly, it aligns with observations indicating
that the mass–size relation for snowflake aggregates typically follows
a consistent scaling law, $m \propto \Dmax^p$ with $p \approx 2$, even
in the presence of depositional growth \citep{Locatelli-Hobbs-1974,
  Mitchell-1990, Leinonen-2021}.

A key limitation of the approach presented here is the absence of
aggregate memory throughout collision events: a new random geometry is
generated for each collision event, with no retention of structural
features from the parent particles. In nature, aggregates likely
preserve aspects of their precursors, such as shape asymmetries or
alignment. Incorporating such memory effects, however, would
significantly complicate the stochastic scheme, as the joint
probability distributions introduced in the previous section would
need to account for the properties of both parent particles resulting
in high-dimensional PDFs. For a Monte Carlo Lagrangian particle model,
in which each super-particle represents an ensemble of real particles,
the neglect of parent-specific memory appears to be a reasonable and
pragmatic simplification. Nonetheless, since \McSnow\ explicitly
tracks the individual properties of parent particles, this information
could, in principle, be used to inform the geometry of the resulting
aggregate in a more physically consistent way.

\subsection{Idealized 1D simulations}

To understand the impact of the stochastic geometry of aggregate
snowflakes on the microstructure of clouds, we perform one-dimensional
simulations using a stationary atmospheric profile similar to that in
\citet{Welss-2024}, Section 5. The temperature profile is defined by a
surface temperature of 273~K and a constant lapse rate of
0.0062~K\,m$^{-1}$, within a domain extending up to 5000~m (case 1)
and 4000~m (case 2). Water vapor, liquid water, and
temperature are prescribed and held constant; no microphysical
processes modify them during the simulation. Particles nucleate
randomly within the uppermost 1500~m of the domain and grow by vapor
deposition at a constant supersaturation with respect to ice of 5\%. Below 1000~m
altitude, they pass through a cloud layer with a liquid water mixing
ratio of 0.1~g/kg, leading to riming similar to a seeder--feeder
cloud. The initial multiplicity is $16\times10^4$, and the vertical grid
spacing of the Eulerian grid is 250~m. These settings result in more
than 2000 super-particles per grid box throughout the column.

For each case, we compare three simulations: one using
the empirical aggregates from \citet[][M96 hereafter]{Mitchell-1996}, one using the
stochastically generated aggregates described in the previous section,
and one using the deterministic mean of the new aggregates. In the
deterministic mean snowflake experiment, all stochastic variability is
removed by setting standard deviations to zero, such that each
snowflake of a given mass and monomer number has identical properties.

\def\mywidth{7.8cm}
\begin{figure}[t]
  \centering
  \begin{minipage}{\mywidth}
    a) number density \\[3mm]
    \includegraphics[width=\textwidth,viewport= 50 150 550 565,page=3,clip=]
                    {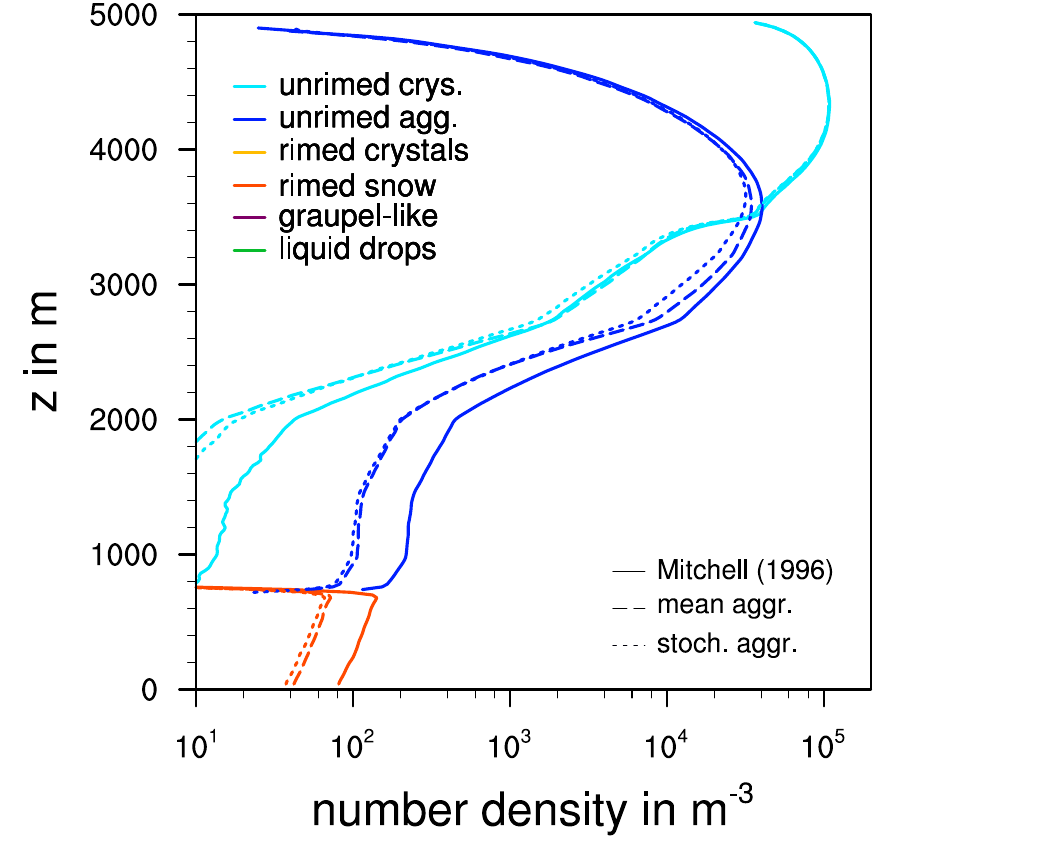}
  \end{minipage}
  \hspace{3mm}
  \begin{minipage}{\mywidth}
    b) mass density \\[3mm]
    \includegraphics[width=\textwidth,viewport= 50 150 550 565,page=4,clip=]
                    {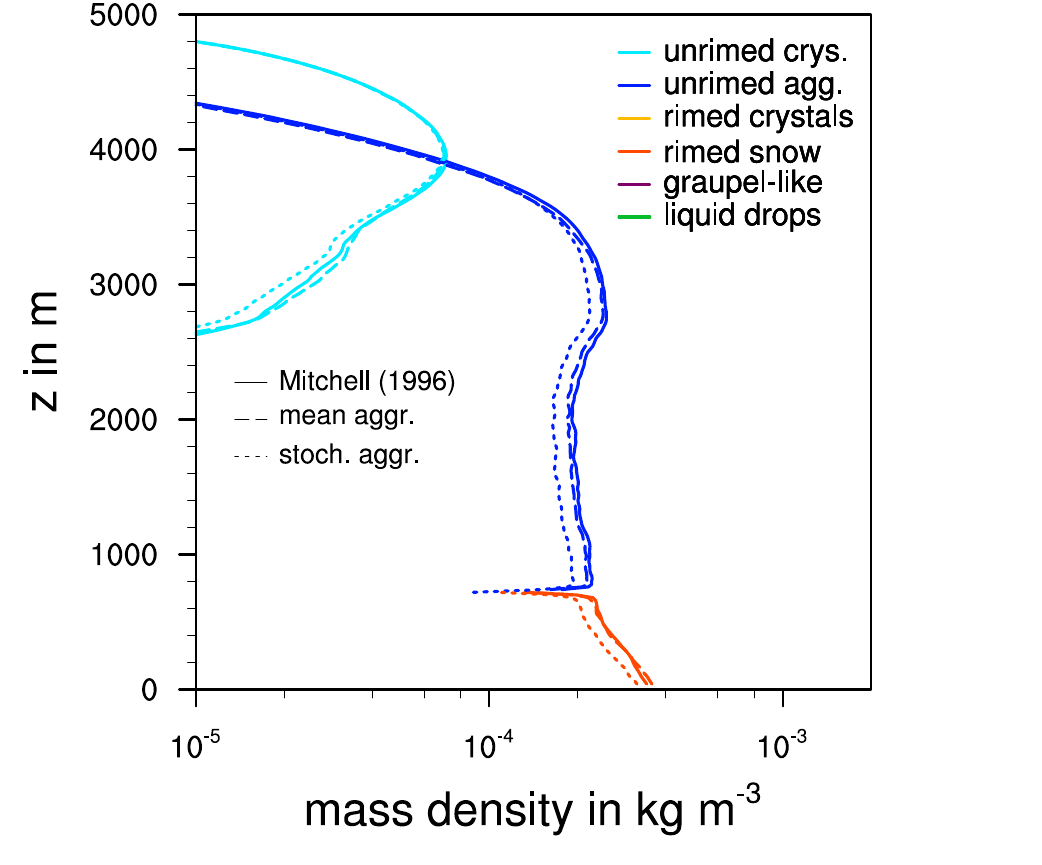}
  \end{minipage}
  \caption{Vertical profiles of number and mass densities for the
    quasi-stationary state of 1D McSnow simulations (case 1). Super-particles
    are categorized as unrimed crystals, unrimed aggregates, rimed
    crystals, rimed aggregates, and liquid drops. The control
    simulation uses empirical aggregate geometry based on M96 (solid
    lines). The simulation using the new stochastically generated
    aggregates is represented by dotted lines. Dashed lines correspond
    to a simulation with the new aggregates, but using zero standard
    deviations, i.e., the deterministic mean aggregate snowflakes.
  \label{fig:hprofiles_case2}}
\end{figure}

Figure~\ref{fig:hprofiles_case2} shows vertical profiles of particle
number and mass density, averaged over the final 3 hours of a 6-hour
simulation. The uppermost region of the domain is dominated by primary
ice crystals growing via vapor deposition. Aggregation rapidly becomes
significant, and by 4000~m, aggregate snowflakes dominate the mass
density. Despite this, primary crystals continue to nucleate and thus
remain dominant in number density above 3500~m. Below 3500~m, where no
nucleation occurs, aggregate snowflakes efficiently deplete primary
ice crystals, leading to a notable drop in their concentration. Only a
few primary particles survive and escape collection by
aggregates. While the mass density of aggregates remains fairly
constant between 3000~m and 1000~m height, the number density
decreases due to aggregation. Below 1000~m, riming increases the mass
density of aggregate snowflakes, which become at least partially
rimed.

The three simulations yield similar results, as these 1D simulations
are strongly constrained by the assumed atmospheric profile. In number density, the
largest difference occurs between the M96 aggregates and the mean
aggregates. In mass density, the stochastic aggregates lead to
significantly lower values below 3500~m height. These differences
emerge in the size and terminal fall velocity of the aggregates, which
in turn influence both mass and number densities.

The fact that number and mass densities of aggregate snowflakes are
slightly reduced when using stochastic sampling suggests that
aggregation between aggregates is enhanced due to increased
variability among the particles. This additional aggregation increases
the average size and terminal velocity of the snowflakes, which in
turn lowers the mass density. A second effect is that more
compact aggregates with $\Dnorm < 1$ have higher terminal fall
velocities, and this is not fully compensated by the slower fall
speeds of elongated aggregates. Hence, on average, stochastic sampling
increases terminal fall velocity and, consequently, reduces mass
density. This nonlinearity of the terminal fall velocity — whereby the mean fall speed of a particle ensemble exceeds the fall speed of the mean geometry — results in an increase in the bulk terminal fall velocity of approximately 5 \%. This behavior arises from the same mechanism described by \citet{Dunnavan-2021}, where variability in snowflake shape leads to enhanced ensemble-mean fall speeds. While Dunnavan reported increases of up to 60 \% in fall speeds of individual particles, our smaller value represents the average effect over a broader particle population.

Overall, the sensitivity to the variability of snowflake geometry, as
represented by the stochastic sampling approach, is weaker than
expected. This indicates that polydispersity arising from variation in
aggregate mass and monomer number dominates the aggregation process,
while random variability in maximum dimension among aggregates with
the same mass and monomer number has only a second-order effect on the
collision rate.

\def\mywidth{7.8cm}
\begin{figure}[t]
  \centering
  \begin{minipage}{\mywidth}
    a) number density \\[3mm]
    \includegraphics[width=\textwidth,viewport= 50 150 550 565,page=3,clip=]
                    {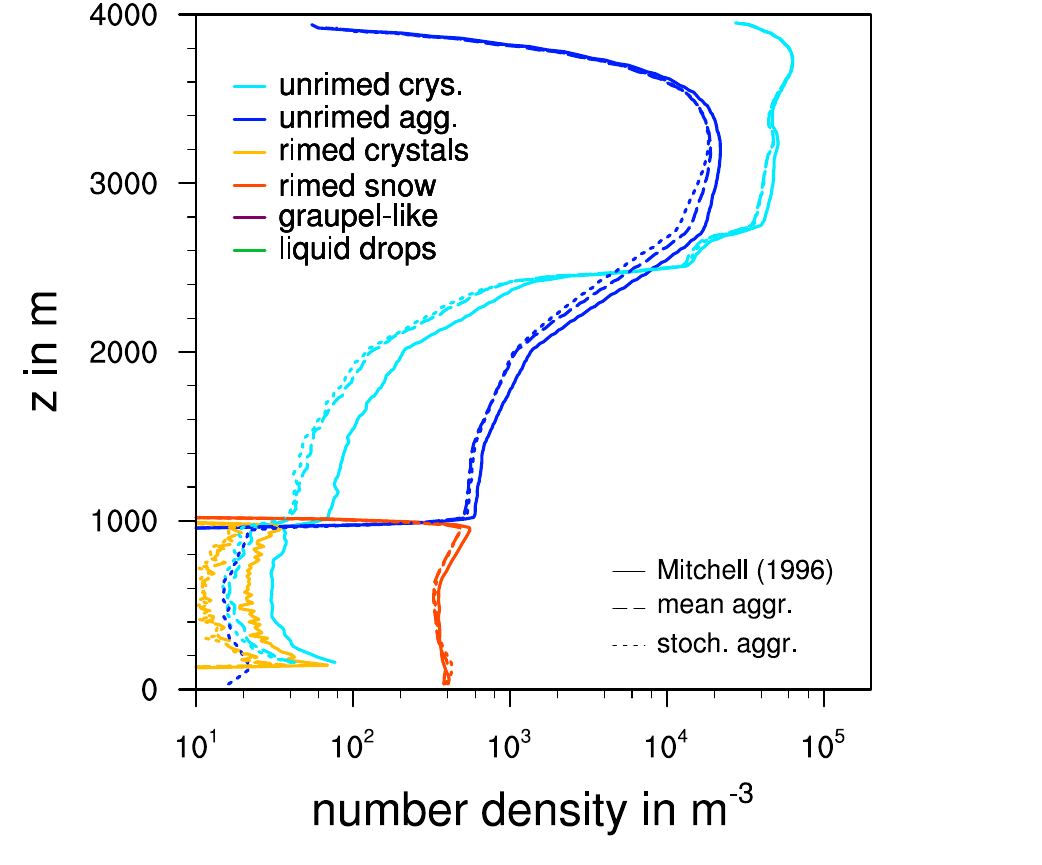}
  \end{minipage}
  \hspace{3mm}
  \begin{minipage}{\mywidth}
    b) mass density \\[3mm]
    \includegraphics[width=\textwidth,viewport= 50 150 550 565,page=4,clip=]
                    {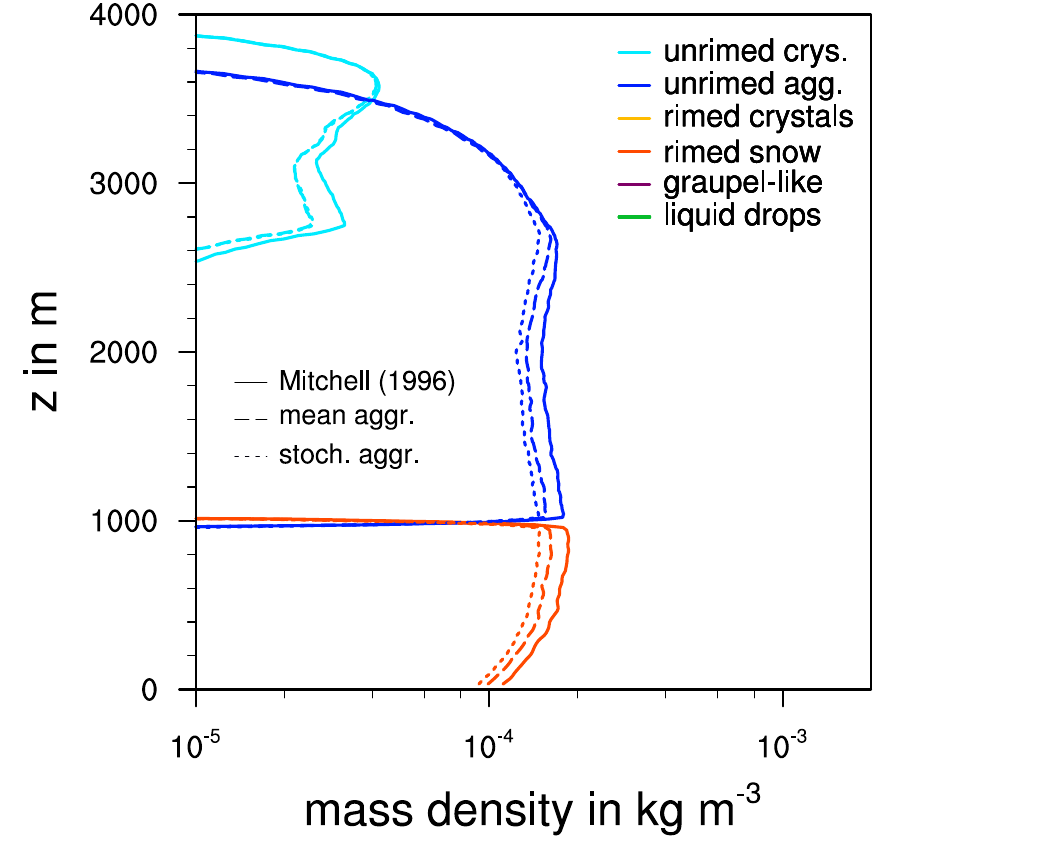}
  \end{minipage}
  \caption{As in Figure~\ref{fig:hprofiles_case2}, but for case 2 with
    a domain top at 4000~m and a nucleation layer in the uppermost
    1500~m.
  \label{fig:hprofile_case1}}
\end{figure}

\def\expname{habit2_xi016_sat5_ncl42_prog_prog}
\def\mywidth{4.9cm}
\begin{figure}[p]
  \centering
  \includegraphics[width=\textwidth,viewport= 700 1200 3600 4500,clip=]
                    {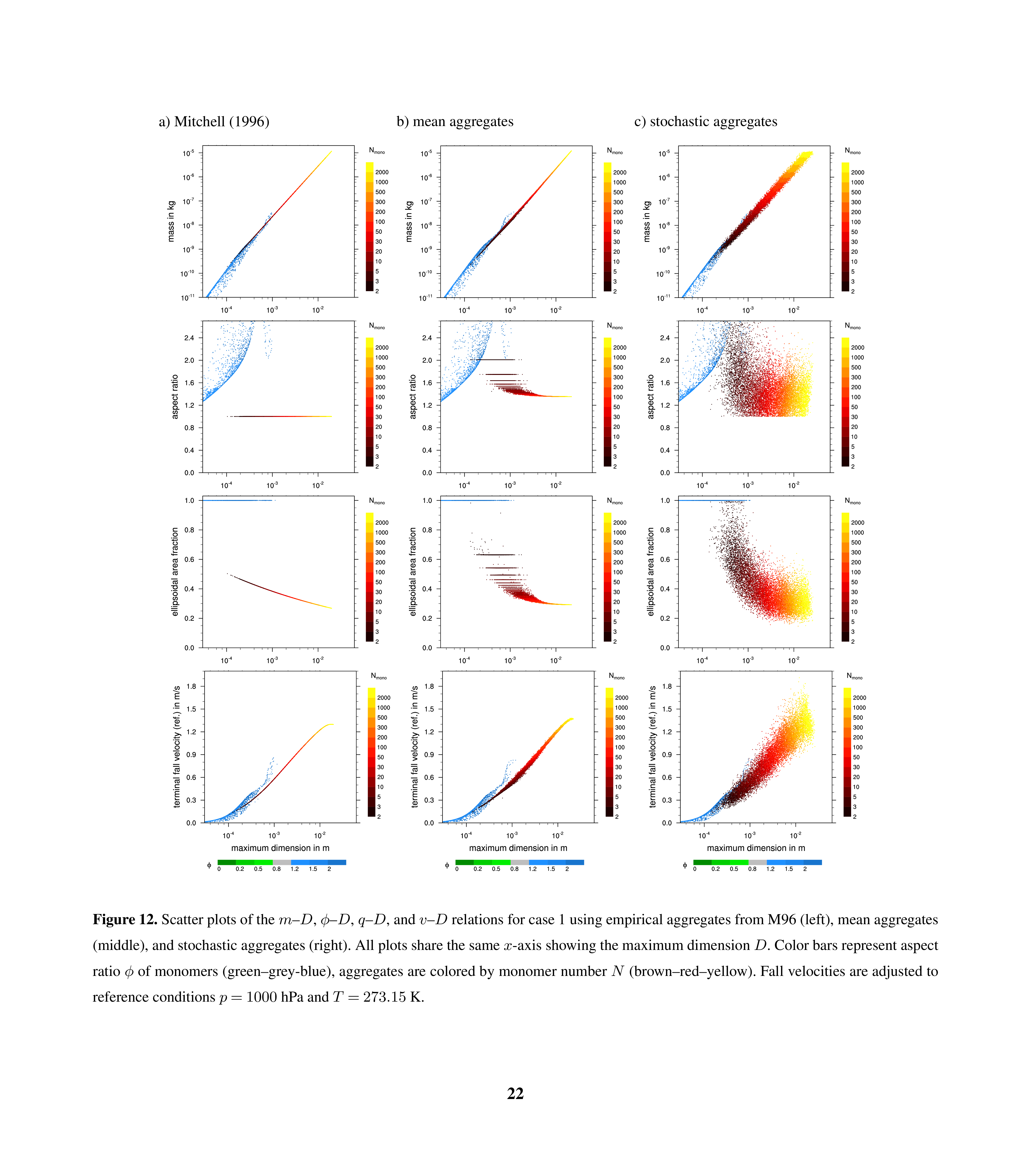}
  \caption{Scatter plots of the $m$--$D$, $\phi$--$D$, $q$--$D$, and
    $v$--$D$ relations for case 1 using empirical aggregates from M96 (left),
    mean aggregates (middle), and stochastic aggregates (right). All
    plots share the same $x$-axis showing the maximum dimension
    $D$. Color bars represent aspect ratio $\phi$ of monomers (green--grey-blue), aggregates are colored by monomer number $N$
    (brown--red--yellow). Fall velocities are adjusted to reference
    conditions $p = 1000$~hPa and $T = 273.15$~K.
  \label{fig:sp_scatter_case1}}
\end{figure}

\def\expname{habit1_xi016_sat5_ncl42_prog_prog}
\def\mywidth{4.9cm}
\begin{figure}[p]
  \centering
  \includegraphics[width=\textwidth,viewport= 700 1000 3600 4500,clip=]
                    {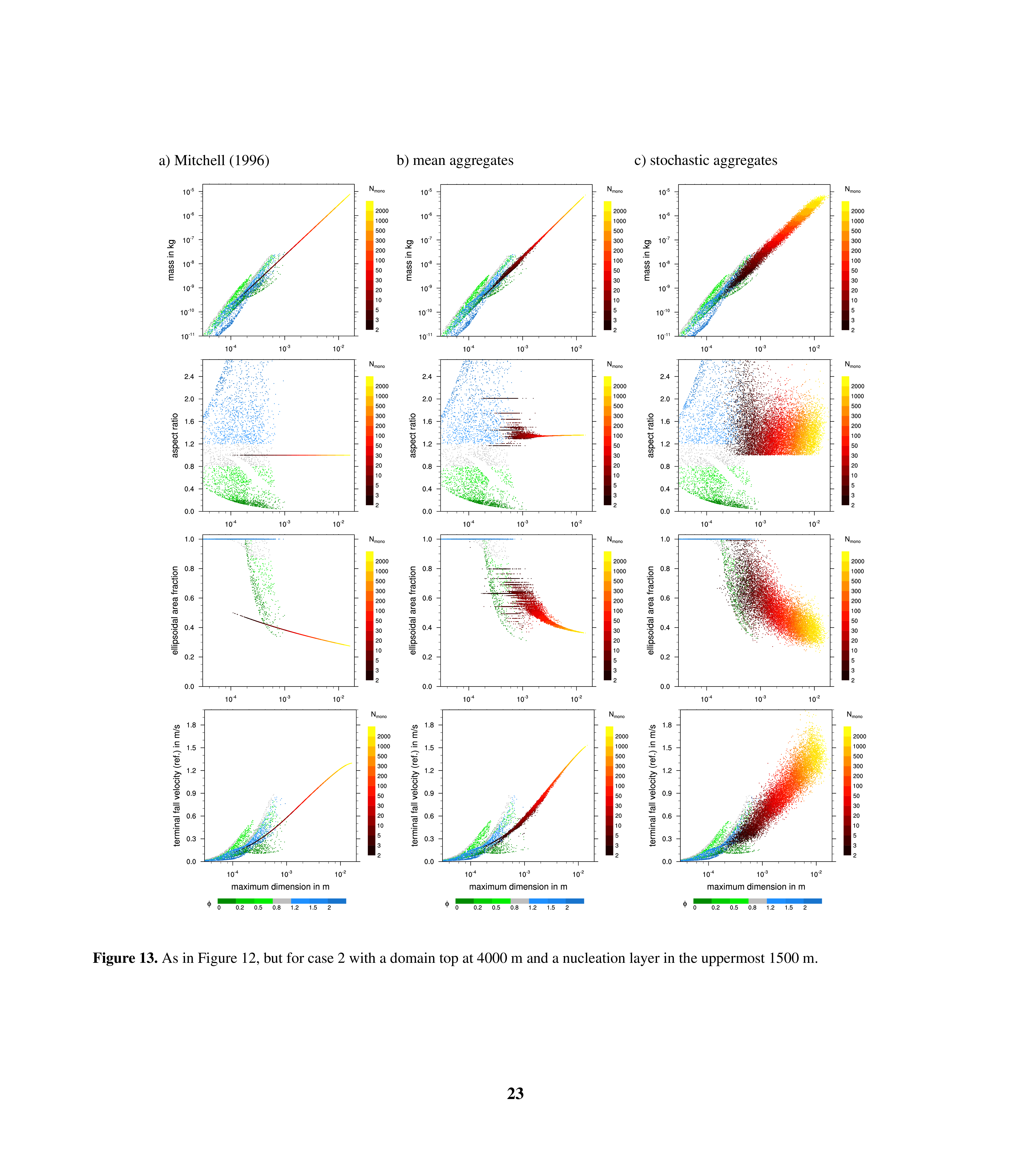}
  \caption{As in Figure~\ref{fig:sp_scatter_case1}, but for case 2 with a domain top at 4000~m and a nucleation layer in the uppermost 1500~m.
  \label{fig:sp_scatter_case2}}
\end{figure}

This is confirmed by case 2, with a domain top at 4000~m. For this
case, which is dominated by dendrites as we will see later, the main
difference in mass density is between M96 and the mean aggregates. The
effect of stochastic sampling is even smaller than in case 1. An
interesting difference between case 1 and case 2 is that the mass
density increases in the riming layer for case 1 but decreases for
case 2. This shows that, depending on the microstructure of
snowflakes, either the mass increase due to riming can dominate (as in
case 1), or the dilution effect due to the increase in terminal fall
velocity (as in case 2).

\def\mywidth{7.0cm}
\begin{figure}[t]
  \centering
    \includegraphics[width=\textwidth]
                    {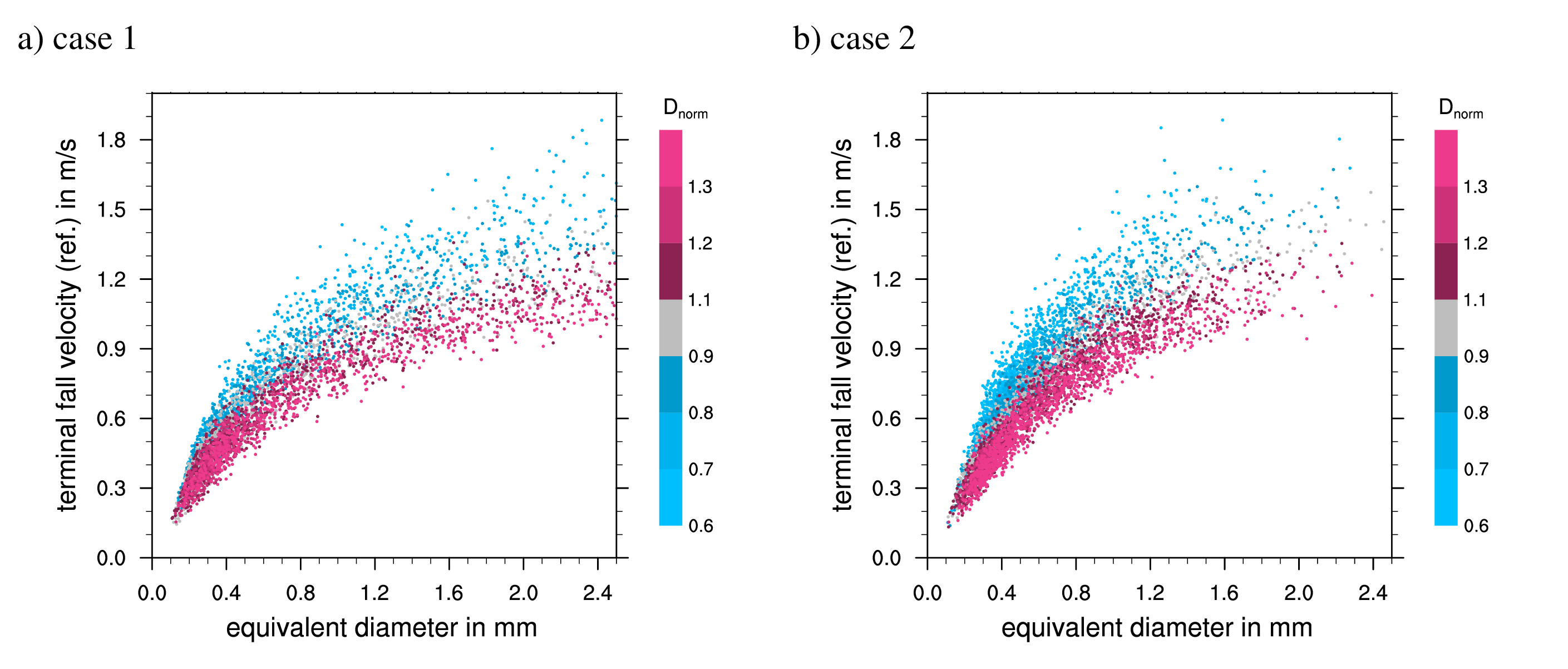}
  \caption{Scatter plot of terminal fall velocity of aggregate
    snowflakes as a function of equivalent diameter for case 1 (left)
    and case 2 (right). Colors indicate the normalized maximum
    dimension $\Dnorm$, with $\Dnorm > 1$ for elongated aggregates and
    $\Dnorm < 1$ for compact aggregates. Velocities are adjusted to
    1000~hPa and 273.15~K.
  \label{fig:velocity_dnorm}}
\end{figure}

To delve deeper into the simulated microstructure of snowflakes and the differences
between case 1 and case 2, Figures~\ref{fig:sp_scatter_case1} and
\ref{fig:sp_scatter_case2} show scatter plots of various ice particle
properties as a function of maximum dimension. Both simulation with stochastic aggregates use
the prognostic approach to aggregate habit described earlier. For the
mass--size relation, the mean aggregates behave similar to the empirical
power law of M96. Using stochastic sampling yields a
variability resembling Figure~1 by construction.

\def\mywidth{8.4cm}
\begin{figure}[t]
  \centering
  \begin{minipage}{\mywidth}
    a) case 1 \\[3mm]
    \def\expname{habit2_xi016_sat5_ncl42_prog_prog}
    \includegraphics[width=\textwidth,viewport= 50 245 585 640,page=1,clip=]
                    {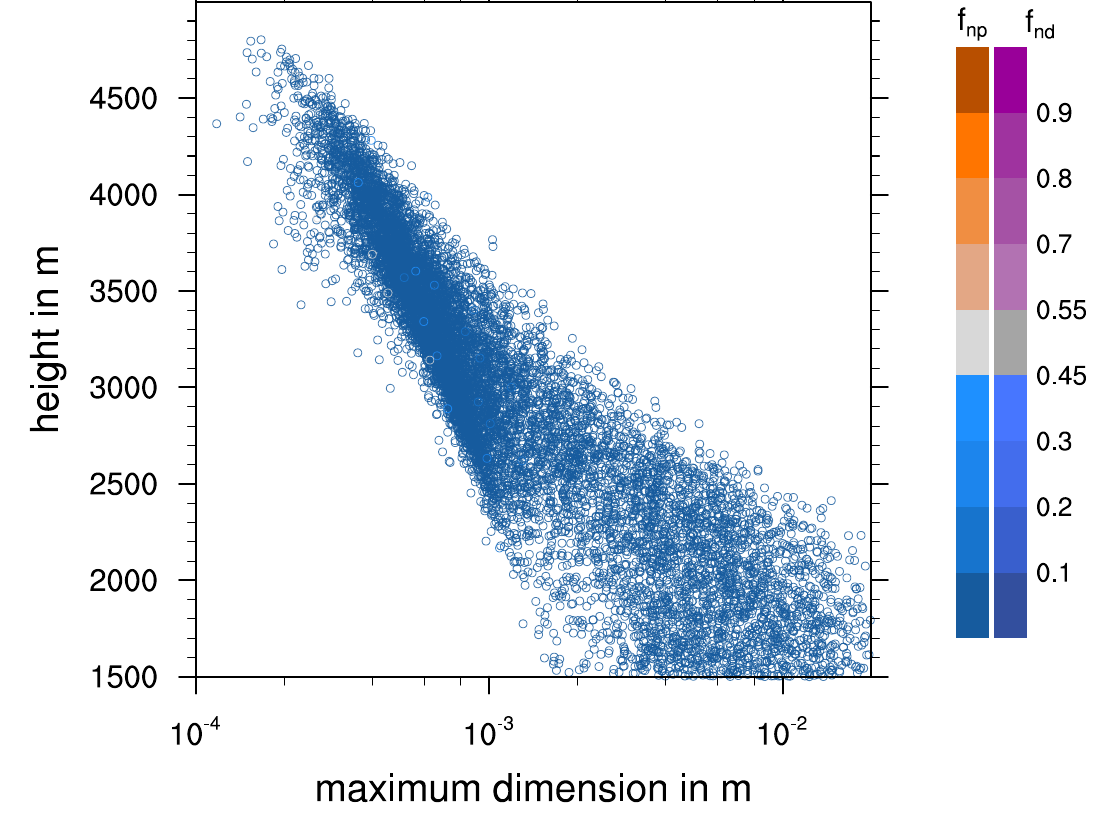}
  \end{minipage}
  \begin{minipage}{\mywidth}
    b) case 2 \\[3mm]
    \def\expname{habit1_xi016_sat5_ncl42_prog_prog}
    \includegraphics[width=\textwidth,viewport= 50 245 585 640,page=1,clip=]
                    {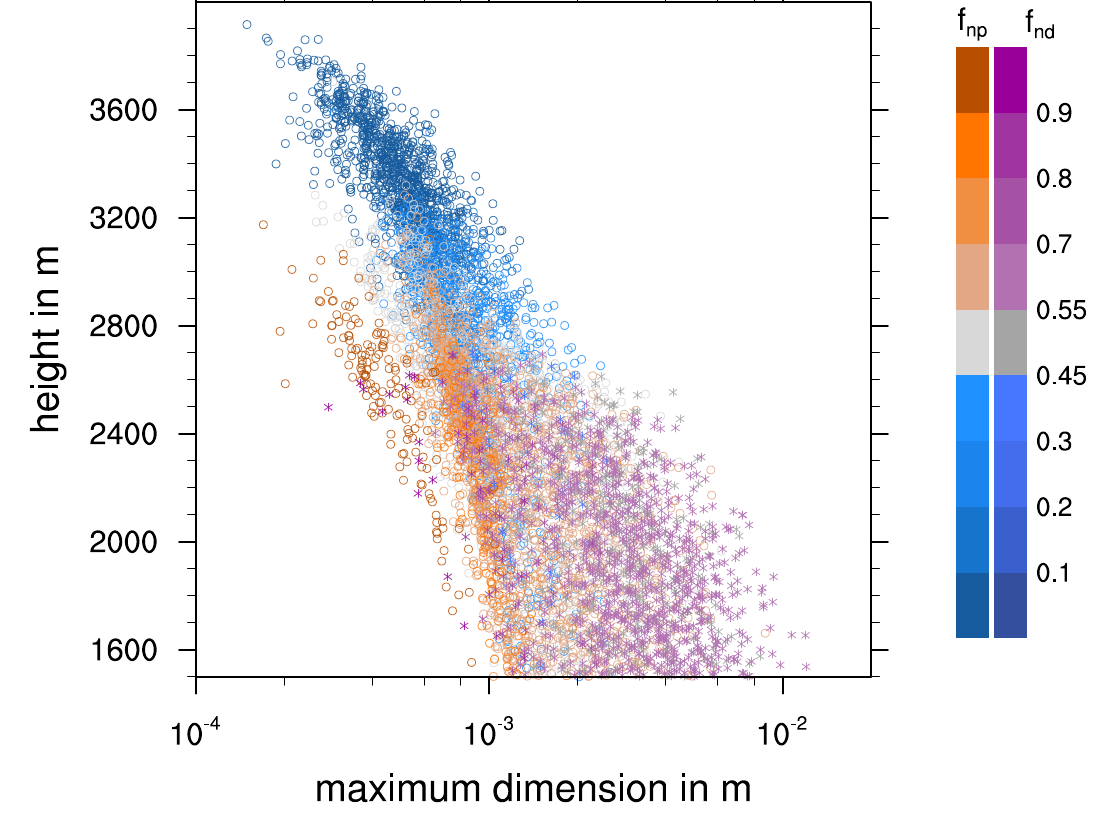}
  \end{minipage}
  \caption{Scatter plot of aggregate snowflake size versus
    height. Colors indicate the number ratio of the binary habit
    mixture, with the dominant monomer habit shown as follows: blue
    for needles, orange for plates, and magenta for dendrites. Marker
    shapes denote mixed habit types: circles for needle–plate
    aggregates and stars for needle–dendrite aggregates. Data points
    have been thinned for clarity.
  \label{fig:aggregate_habits}}
\end{figure}

M96 does not provide an aspect ratio for aggregates,
so in McSnow we assume $\phi = 1$ for M96 aggregates. The new
aggregates are slightly prolate with $\phi = \phi_h^{-1} \approx 1.3$ in the
deterministic mean (see Figure~\ref{fig:aspect-ratio}). The
ellipsoidal area ratio $q$ decreases with size. M96 and our new
deterministic aggregates behave similarly. That M96 has a lower $q$
for small aggregates partly results from our assumption of
$\phi=1$. Horizontal lines for mean aggregates in
Figure~\ref{fig:sp_scatter_case1}b reflect that $\phi$ and $q$ are
functions of monomer number $N$ only. Case 1 is dominated by needle
monomers, shown as blue markers for primary crystals. Case 2 shows a
wider range of habits from prolate to oblate, yielding slightly larger
variability in aggregate properties.

The stochastic variability is especially large for aspect ratio $\phi$
and area fraction $q$. The new aggregates produce fall velocities
similar to M96, with realistic spread, consistent with in-situ
observations. This variability results directly from differences in
$\phi$ and $q$. For primary crystals, prolate particles fall
significantly faster than oblate ones of the same mass. Plotting
terminal fall velocity as function of maximum dimension gives a more
complicated behaviour and oblate particles can indeed fall faster than
prolate crystals of the same size. Large aggregate snowflakes show
terminal velocities up to 1.5~m/s in the mean aggregate case, rising
to 1.8~m/s with stochastic variability. For case 2, the terminal fall
velocity of the new aggregate snowflakes with $\Dmax > 1$ mm is
significantly higher than with M96 aggregates for this
case. Figure~\ref{fig:velocity_dnorm} confirms that compact aggregates
($\Dnorm < 1$) fall faster than elongated ones ($\Dnorm > 1$).

Figure~\ref{fig:aggregate_habits} shows the dominant habits in the
simulated aggregates. Case 1 is consistently dominated by aggregates
of needles at all altitudes. In case 2, aggregates of needles dominate
above 3000~m, with a transition to plate- and dendrite-dominated
aggregates below. Near the surface a rich mix of aggregate habits is
found with large aggregates dominated by dendrites. Validating these
predictions with modern video disdrometers requires a more realistic
model setup and is left for future work.

\section{Simulating radar signatures of stochastic aggregates}

\subsection{A radar forward modeling framework for McSnow}

Forward models, especially in the microwave region, are key components for atmospheric model evaluation and data assimilation. For these forward models, the optical properties of particles simulated in the atmospheric model need to be known. To this end, the discrete-dipole approximation (DDA; \cite{draine1994discrete}) is often employed, as it provides an accurate approximation of the scattering properties of individual particles with arbitrary shapes. In \cite{vonTerzi-2025}, a database containing the scattering properties of 3,077 individual ice crystals was developed, which is particularly suitable for the model output of McSnow. For the present study, this database was extended to include 7,010,688 aggregates. The three-dimensional representations of these aggregates were generated as described in Section~\ref{sec:agg_modelling}.

The Amsterdam DDA (ADDA; \cite{yurkin2007discrete,yurkin2011discrete}) code is used to calculate the optical properties of these particles at common cloud and precipitation radar frequencies, including 5.6, 9.6, 35.6, and 94 GHz. All particles are assumed to be aligned with their largest axis horizontally. For each particle, only a single radar viewing angle is simulated, so that at all frequencies for a given particle at one elevation and one azimuth angle are computed. This approach differs from most previous studies, where each individual particle is typically simulated at multiple orientations \citep[e.g.][]{Lu-2016,brath2020microwave,vonTerzi-2025}. For example, in \cite{vonTerzi-2025}, each particle is simulated at 37 elevations and 16 azimuth orientations. Such an approach is often chosen to reduce the number of distinct particles that need to be generated, as especially for aggregates, the scattering properties can vary significantly with viewing angle.

In contrast, our study leverages the large aggregate database to provide substantial variability in scattering properties even when only one radar viewing angle per particle is considered. In total, we calculated scattering properties for 7,010,688 particles. For comparison, \cite{vonTerzi-2025} included 3,077 ice crystals simulated at 37 elevations and 16 azimuth orientations, corresponding to 1,821,584 individual calculations, approximately four times fewer than the number of aggregates used here. The major advantage of calculating the scattering properties for such a large number of particles is the wide variability in physical properties (e.g., $D_{\max}$, mass, aspect ratio) that can be simulated. While the total number of scattering calculations in our study is only about four times larger than that of the crystals used in \cite{vonTerzi-2025}, the total number of different physical properties covered is approximately 2,300 times larger. This makes it particularly suitable for models with widely varying physical properties, such as the stochastic aggregates introduced here.

For each particle, the amplitude and Mueller matrix entries at 0° and 180° scattering directions are calculated. From these matrix entries, the single-particle reflectivity at horizontal (Ze$_{\rm H}$) and vertical (Ze$_{\rm V}$) polarizations, the extinction cross sections at horizontal (c$_{\rm ext,h}$) and vertical (c$_{\rm ext,v}$) polarizations, as well as the specific differential phase shift KDP, are derived using well-known equations \citep[e.g.,][]{vonTerzi-2025}. These values are stored in a lookup table (LUT), which is then interfaced to the forward operator McRadar as described in detail in \cite{vonTerzi-2025}. 

In \cite{vonTerzi-2025}, the focus was on super-particles, where an azimuthal average of the scattering properties was calculated for each particle. These azimuthally averaged scattering properties were then assigned uniformly to each super-particle. This means that all the real particles represented by a single super-particle shared the same averaged scattering properties, simplifying the calculation but potentially smoothing out orientation variability.

In contrast, our new implementation operates at the level of real particles represented by each super-particle and samples the scattering properties based on a single radar viewing angle per particle (so one elevation and one azimuth angle are sampled per particle), assuming the particles fall with their largest dimension oriented horizontally. Since only one radar viewing angle is considered for each real particle, azimuthal averaging of scattering properties cannot be performed as before. Instead, the best-fitting scattering properties of individual particles are assigned to each simulated real particle, allowing for slight variations in scattering properties among the real particles associated with each super-particle.

This approach is analogous to the previous averaging method in the following way: if a super-particle has a multiplicity of 16, matching the number of azimuth directions used in \cite{vonTerzi-2025}, one could theoretically assign the scattering properties from each azimuth angle to a corresponding real particle within that super-particle. The total reflectivity ($Ze$) would then be obtained by summing the contributions from all real particles. Alternatively, the older approach averaged over all azimuth angles first and assigned this average to the entire super-particle, then multiplied by the multiplicity to obtain the total reflectivity. Both methods yield equivalent results in principle.

Although in our case the multiplicity does not typically equal the number of azimuth directions probed, the very large number of particles considered ensures that the overall statistical behavior of the scattering properties remains comparable.

The LUT contains information on particle mass, size, habit, and elevation, while McSnow provides mass, size, habit, and the desired radar setup provides the radar elevation. The goal is to retrieve particles whose mass, size, habit, and elevation correspond to those in the LUT. Here, \textit{habit} refers to the types of monomers in the aggregate (i.e., needles, dendrites, and plates) and their ratios (see Section~\ref{sec:apl_mcsnow} for more details). The selection is performed using a K-Dimensional Tree (K-D Tree), where for each super-particle, candidates from the lookup table are identified based on target mass, size, habit, and elevation. A variability of 5\% in mass and $D_{\max}$, 50\% in monomer ratio, and 5° in elevation is allowed. The variability in elevation allows for a flexible treatment of the particle canting angle, simulating wobbling effects due to, e.g., turbulence in the overall fall behavior of the particles. Out of all candidates, $N = \min(n_{\rm cand}, \xi)$ are randomly selected, where $n_{\rm cand}$ is the number of found candidates and $\xi$ is the multiplicity of the super-particle. These are weighted by $wgt = \xi / n_{\rm cand}$, ensuring that $N \cdot wgt = \xi$. This weighing needs to be done, as for super-particles with large $\xi$ and small variabilities in mass, $D_{\max}$ and monomer ratio, not enough candidates can be found to select one candidate for each real particle. The scattering properties of the super-particle are then obtained as the weighted sums of the scattering properties of all selected candidates.

\subsection{Simulated radar signatures of case 2}
\begin{figure}[t]
    \centering
    \includegraphics[width=\linewidth]{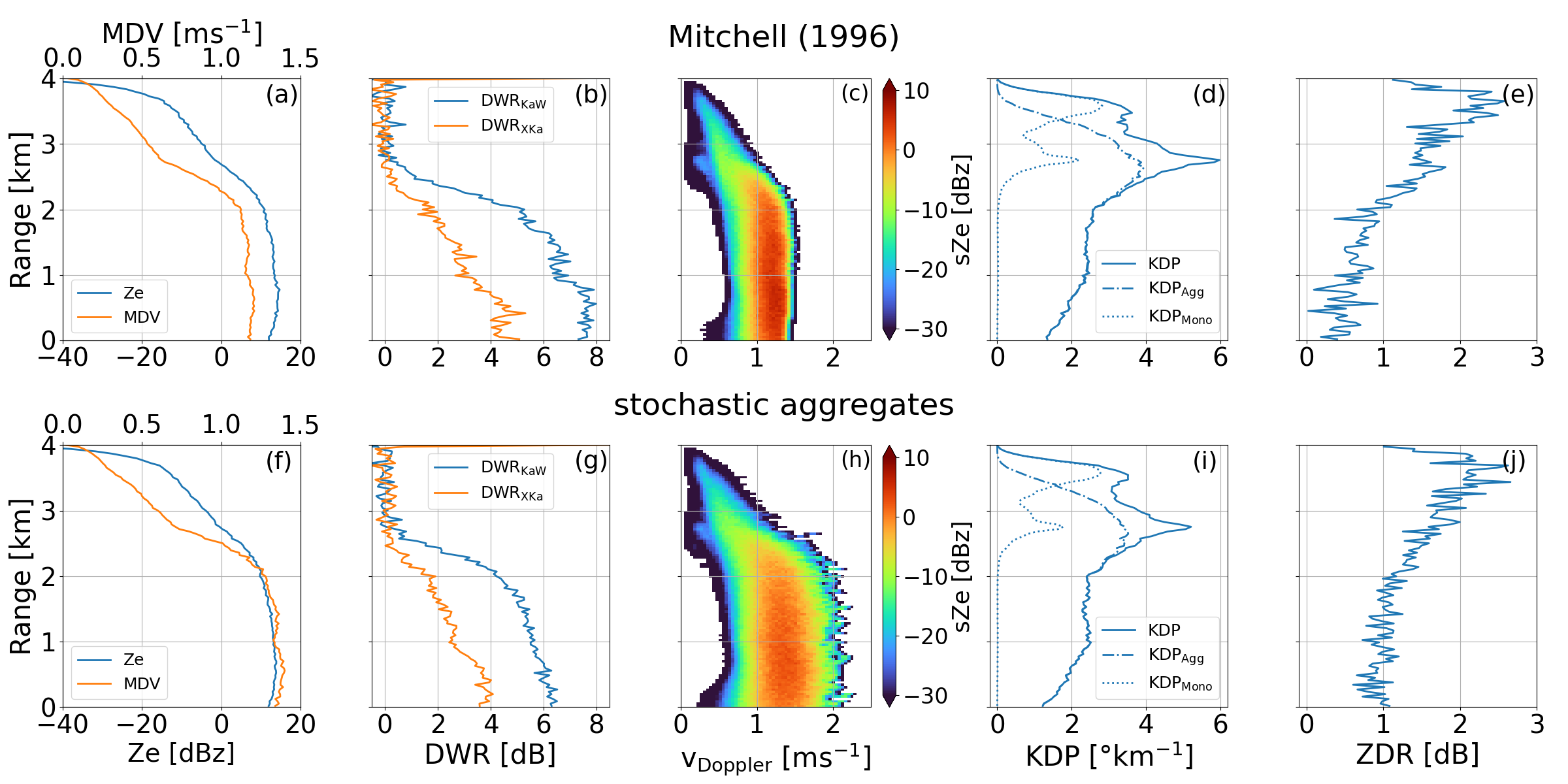}
    \caption{Radar forward simulations of case 2 using empirical aggregates from M96 (first row) and stochastic aggregates (second row). The first column (panels (a) and (f)) shows the radar reflectivity (Ze) and mean Doppler velocity (MDV) at Ka-band (35.6 GHz) and 90° elevation (zenith view). The second column (panels (b) and (g)) shows the reflectivity differences (in dB, commonly denoted as dual-wavelength ratio) at zenith between Ka- and W-band (94 GHz) and X (9.6 GHz) and Ka-Band. The third column (panels (c) and (h)) shows the zenith spectral reflectivity (sZe) at Ka-Band, the fourth column (panels (d) and (i)) shows the specific differential phase shift KDP at W-Band and 30° elevation, as well as the contributions of aggregates and monomers to the total KDP. The fifth column (panels (e) and (j)) shows the differential reflectivity ZDR at 30° elevation and W-Band.}
    \label{fig:case_study_forward_sim}
\end{figure}

To compare the radar signatures of the M96 aggregates and the stochastic aggregates, case 2 was forward simulated at 9.6 (X-band), 35.6 (Ka-band), and 94 (W-band) GHz, and at 30° and 90° elevation. Figure~\ref{fig:case_study_forward_sim} shows the main radar moments: radar reflectivity factor (Ze), mean Doppler velocity (MDV), the dual-wavelength ratios (reflectivity difference in dB) between Ka- and W-Band DWR$_{\rm KaW}$ and between X- and Ka-Band DWR$_{\rm XKa}$, the polarimetric variables specific differential phase shift (KDP) and differential reflectivity (ZDR), as well as the spectral reflectivity (sZe). Positive Doppler velocities indicate motions towards the radar.

The case study displays typical features often observed in ice cloud measurements \citep[][and references therein]{von2022ice}. Aggregation increases both Ze and DWR, at Ka- and W-band as well as at X- and Ka-band, especially towards the ground. The strongest increase in DWR occurs at temperatures between –20 °C and –10 °C, corresponding to the dendritic growth layer (DGL), which is known for high aggregation efficiencies \cite{Connolly2012}. In the same region, the Doppler spectra show increasing sZe, consistent with particle growth. Interestingly, at 2.75 km (–17 °C), a secondary, slowly falling mode appears in the spectrum. This new mode arises from freshly formed, small, plate-like ice crystals at that temperature. Although aggregation acts as a sink for such crystals, collision efficiency is strongly size-dependent; the smallest crystals have the lowest chance of colliding with aggregates. As a result, a distinct mode of small ice crystals develops. Closer to the ground, these crystals grow and fall faster, while no new crystals are produced. Consequently, the secondary mode merges with the main one and becomes indistinguishable, even though small crystals remain present down to near-surface levels.

KDP shows a pronounced peak at 2.75 km, coinciding with the appearance of the secondary Doppler spectra mode. KDP typically reflects increases in particle size and concentration, and here it indicates large numbers and sizes of ice particles at this height. The peak is caused by both aggregates and monomers, though aggregates dominate the signal. As aggregation continues, the total number concentration decreases, and KDP correspondingly declines towards the ground.

ZDR, in contrast, peaks at 3.5 km, slightly higher than the main KDP maximum. The peak in ZDR coincides with the largest KDP of monomers. However, towards the ground, aggregates dominate KDP in this case, off-setting the main KDP peak and the ZDR peak. This vertical separation of ZDR and KDP has been reported in many earlier studies \cite[e.g.][]{von2022ice,Moisseev2015}, although the underlying cause remains debated.

When comparing the two experiments, the overall radar signatures appear similar, but important differences emerge. The most evident difference is seen in the Doppler spectra: it is much broader for the stochastic aggregates, reflecting the large variability in their physical properties, especially in terminal fall velocity (cf.~Fig.~\ref{fig:sp_scatter_case2}). The fast-falling edge extends to 2 m~s$^{-1}$ for the stochastic aggregates, compared to 1.5 m~s$^{-1}$ for the M96 aggregates. Consequently, MDV is larger in the stochastic case (1.4~m~s$^{-1}$ versus 1.2~m~s$^{-1}$). Cloud radar observations in mid-latitude frontal clouds commonly show a Doppler spectrum width around 2 m/s above the melting layer (see e.g. Figs. 3 and 5 of \citet{von2022ice}) similar to the McSnow simulation with stochastic aggregates. 

The mean aggregate size is similar between the two experiments. Surprisingly, DWR$_{\rm KaW}$ is smaller for the stochastic aggregates (a maximum of 6.1~dB in contrast to 7.7~dB), while DWR$_{\rm XKa}$ remains the same. For a single particle type (e.g. dendritic aggregates with a uniform structure), a smaller DWR usually implies smaller aggregate size. However, the internal mass distribution also matters: denser particles can yield larger DWR$_{\rm KaW}$ but similar DWR$_{\rm XKa}$ compared to less dense ones \citep{Mason2019,kneifel2015observed}. Here, the habit of the aggregates and therefore their internal structure is of course different between the two experiments, the smaller DWR$_{\rm KaW}$ might indicate that the stochastic aggregates are less dense. However, in addition, the aggregate size distribution itself affects the relationships between DWR$_{\rm XKa}$ and DWR$_{\rm KaW}$ \citep{Mason2019}. Looking at the size distributions for this case study, it becomes evident that the size distribution of the stochastic aggregates is wider, encompassing larger and smaller particles than the M96 experiment. Therefore, a combination of internal structure and differing size distributions between the stochastic and M96 aggregates likely explains the smaller DWR$_{\rm KaW}$. 

The polarimetric signals are similar, though the KDP peak is slightly weaker in the stochastic case (5~°~km$^{-1}$ for stochastic aggregates compared to 6~°~km$^{-1}$ for the M96 aggregates) because both aggregates and crystals produce lower KDP values. Also, ZDR at heights smaller than 2~km is larger, saturating at 1~dB for the stochastic aggregates compared to 0.5~dB for the M96 aggregates. Looking at spectrally resolved ZDR (not shown), the larger ZDR is caused by the large aggregates populating the fast-falling side of the spectrum. These aggregates are not present in the M96 simulation.

Overall, the forward simulations reveal a pronounced effect of the stochastic formulation of aggregate snowflakes in McSnow on the Doppler spectra, multi-frequency and polarimetric radar quantities. A validation with actual radar observations is part of an ongoing research project, but beyond the scope of the current study.



\section{Summary and Conclusions}

This study presents a novel stochastic framework to describe the
geometry of aggregate snowflakes within Monte Carlo Lagrangian
particle models like \McSnow. Unlike traditional deterministic or
empirically parameterized geometric representations of snowflakes,
this approach leverages joint probability distributions derived from
extensive datasets of simulated aggregates, generated using a detailed
geometric aggregation model. These distributions are used to sample
key microphysical and geometric variables: maximum dimension, aspect
ratio, and area ratio. This method aims to capture the inherent
variability and complexity observed in natural snowflakes, providing a
more physically consistent description of their microstructure.

Another key improvement of this framework is its ability to model mixtures
of monomer habits through a continuously defined metric of monomer
habits of aggregates of mixtures. This enables flexible interpolation
of geometric properties across populations dominated by different
monomer types, such as plates, needles, and dendrites, thereby
capturing the wide diversity in snowflake morphology observed in
clouds. In the Lagrangian particle model, maximum dimension is treated
as a pseudo-prognostic variable, dynamically updated during collision
and depositional growth, while other geometric parameters like aspect
ratio are assumed constant for aggregates during depositional
growth. Some of these simplifications reflect the limits of our
current knowledge of snowflake growth.

Idealized one-dimensional simulations of snowflake aggregation
demonstrate that the inclusion of stochastic geometric variability
enhances aggregational growth by increasing variability in particle
terminal velocities. This leads to marginally larger aggregate sizes
and somewhat reduced number and mass concentrations relative to models
using classical empirical aggregate representations. These effects
arise because geometric variability broadens the distribution of
particle fall speeds, increasing the collision kernel and,
consequently, the collision rate. Furthermore, case studies
contrasting dendrite- and needle-dominated situations highlight the
model’s capability to represent distinct microstructural regimes,
emphasizing its potential for improving microphysical realism in
cloud-scale simulations.

Validation against in-situ observational datasets, particularly those
collected by modern video disdrometers and imaging probes, is
essential to rigorously assess the model’s accuracy in representing
snowflake microstructure and to quantify its implications for cloud
microphysics and precipitation processes but is beyond the scope of
the current study. Such validation would provide critical feedback for
refining model assumptions, parameterizations, and identify missing
physics.

Complementary to in-situ observations, we have shown that a validation
of aggregate snowflake properties is possible through forward
simulations of multi-frequency and polarimetric radar variables. Using
a lookup table derived from DDA calculations of millions of stochastic
aggregates, we simulated radar observables such as reflectivity,
Doppler velocity, dual-wavelength ratios, and polarimetric moments
across multiple frequencies and viewing geometries. These forward
simulations reveal that the stochastic aggregate model reproduces key
features seen in observed ice cloud profiles, while also producing
broader Doppler spectra and more variable dual-wavelength signals than
traditional empirical aggregate models. This highlights the diagnostic
value of radar forward modeling for assessing aggregate variability
and evaluating microphysical model performance against remote sensing
data.

The framework currently assumes that each collision generates a new
aggregate geometry independently, without memory of the parent
particles' structure. However, since \McSnow\ tracks detailed parent
properties during collisions, it provides a natural foundation for
incorporating structural memory in future models. Such memory-aware
approaches could improve realism, though at the cost of increased
complexity. Machine learning techniques may offer a path to reduce
this cost by emulating high-dimensional memory effects.

Another open question is whether sticking efficiency depends on
crystal habit. While dendritic particles are often assumed to
aggregate more readily than plates due to their branched structure,
direct observational evidence is scarce. Implementing habit-dependent
sticking in \McSnow\ would require empirical constraints from field or
lab data.

More broadly, several key physical uncertainties remain: How long do
aggregates take to settle into stable structures? Do branches break
and re-freeze, leading to compaction or secondary ice production? How
does vapor deposition reshape aggregate geometry, and how does this
depend on monomer habit and environmental conditions? Despite decades
of research, many of these fundamental processes remain poorly
understood, highlighting the need for continued observational and
modeling efforts.

In conclusion, the stochastic aggregate snowflake framework developed
here constitutes a significant advance toward capturing the natural
variability and complexity of ice particle geometry in cloud
microphysics models. By representing realistic geometric variability
and enabling flexible treatment of monomer habit mixtures, this
approach enhances predictive capability for simulating cloud
microphysical processes and precipitation formation. Moreover, the
framework provides a robust foundation for incorporating more
sophisticated physical effects in the future, ultimately contributing
to improved representation of ice microphysics in weather and climate
models.

\section*{Acknowledgements}

The study contributes to and is partly funded by the project FRAGILE
of the SPP 2115 Fusion of Radar Polarimetry and Numerical
Atmospheric Modelling Towards an Improved Understanding of Cloud and
Precipitation Processes funded by the Deutsche
Forschungsgemeinschaft (DFG 492234709). The authors would further
like to thank Davide Ori for discussions about the performance of
DDA calculations and orientations of ice particles.



\bibliographystyle{copernicus}
\bibliography{snowflake.bib}

@string{ACP     = {Atmos.~Chem.~Phys.}}

@string{AMT     = {Atmos.~Meas.~Tech.}}

@string{GRL     = {Geophys.~Res.~Lett.}}

@string{GMD     = {Geosci.~Model~Dev.}}

@string{JAS     = {J.~Atmos.~Sci.}}

@string{JAM     = {J.~Appl.~Met.}}

@string{JGR     = {J.~Geophys.~Res.}}

@string{QJ      = {Quart.~J.~Roy.~Met.~Soc.}}

@string{JAMES   = {J.~Adv. in Modeling Earth Systems}}

@Article{vonTerzi-2025,
AUTHOR = {von Terzi, L. and Ori, D. and Kneifel, S.},
TITLE = {A Microwave Scattering Database of Oriented Ice and Snow Particles: Supporting Habit-Dependent Growth Models and Radar Applications (McRadar 1.0.0) },
JOURNAL = {EGUsphere [preprint]},
YEAR = {2025},
URL = {https://egusphere.copernicus.org/preprints/2025/egusphere-2025-3910/},
DOI = {10.5194/egusphere-2025-3910}
}

@Article{Lu-2016,
AUTHOR = {Lu, Y. and Jiang, Z. and Aydin, K. and Verlinde, J. and Clothiaux, E. E. and Botta, G.},
TITLE = {A polarimetric scattering database for non-spherical ice particles at microwave wavelengths},
JOURNAL = AMT,
VOLUME = {9},
YEAR = {2016},
NUMBER = {10},
PAGES = {5119--5134},
URL = {https://amt.copernicus.org/articles/9/5119/2016/},
DOI = {10.5194/amt-9-5119-2016}
}

@article{brath2020microwave,
  title={Microwave and submillimeter wave scattering of oriented ice particles},
  author={Brath, Manfred and Ekelund, Robin and Eriksson, Patrick and Lemke, Oliver and Buehler, Stefan A},
  journal=AMT,
  volume={13},
  number={5},
  pages={2309--2333},
  year={2020},
  publisher={Copernicus GmbH},
  doi={10.5194/amt-13-2309-2020}
}

@article{kneifel2015observed,
  title={Observed relations between snowfall microphysics and triple-frequency radar measurements},
  author={Kneifel, S. and von Lerber, A. and Tiira, J. and Moisseev, D. and Kollias, P. and Leinonen, J.},
  journal=JGR,
  volume={120},
  number={12},
  pages={6034--6055},
  year={2015},
  publisher={Wiley Online Library},
  doi={https://doi.org/10.1002/2015JD023156}
}

@article{Mason2019,
AUTHOR = {Mason, S. L. and Hogan, R. J. and Westbrook, C. D. and Kneifel, S. and Moisseev, D. and von Terzi, L.},
TITLE = {The importance of particle size distribution and internal structure for triple-frequency radar retrievals of the morphology of snow},
JOURNAL = AMT,
VOLUME = {12},
YEAR = {2019},
NUMBER = {9},
PAGES = {4993--5018},
DOI = {10.5194/amt-12-4993-2019}
}

@article{Connolly2012,
author = {Connolly, P. J. and Emersic, C. and Field, P. R.},
doi = {10.5194/acp-12-2055-2012},
journal = ACP,
number = {4},
pages = {2055--2076},
title = {{A laboratory investigation into the aggregation efficiency of small ice crystals}},
volume = {12},
year = {2012}
}

@article{Moisseev2015,
author = {Moisseev, D. N. and Lautaportti, S. and Tyynela, J. and Lim, S.},
doi = {10.1002/2015JD023884},
journal = JGR,
number = {24},
pages = {12,644--12,665},
title = {Dual-polarization radar signatures in snowstorms: Role of snowflake aggregation},
volume = {120},
year = {2015}
}

@article{von2022ice,
  title={Ice microphysical processes in the dendritic growth layer: a statistical analysis combining multi-frequency and polarimetric Doppler cloud radar observations},
  author={von Terzi, L. and Dias Neto, J. and Ori, D. and Myagkov, A. and Kneifel, S.},
  journal=ACP,
  volume={22},
  pages={11795-11821},
  year={2022},
  publisher={Copernicus GmbH},
  doi={10.5194/acp-22-11795-2022}
}

@article{draine1994discrete,
  title={Discrete-dipole approximation for scattering calculations},
  author={Draine, Bruce T and Flatau, Piotr J},
  journal={Journal of the Optical Society of America A},
  volume={11},
  number={4},
  pages={1491--1499},
  year={1994},
  publisher={Optical Society of America},
  doi={10.1364/JOSAA.11.001491}
}

@article{yurkin2007discrete,
  title={The discrete dipole approximation: an overview and recent developments},
  author={Yurkin, Maxim A and Hoekstra, Alfons G},
  journal={J. Quant. Spectrosc. Radiat. Transf.},
  volume={106},
  number={1-3},
  pages={558--589},
  year={2007},
  publisher={Elsevier},
  doi={10.1016/j.jqsrt.2007.01.034}
}

@article{yurkin2011discrete,
  title={The discrete-dipole-approximation code ADDA: capabilities and known limitations},
  author={Yurkin, Maxim A and Hoekstra, Alfons G},
  journal={J. Quant. Spectrosc. Radiat. Transf.},
  volume={112},
  number={13},
  pages={2234--2247},
  year={2011},
  publisher={Elsevier},
  doi={10.1016/j.jqsrt.2011.01.031}
}

@article{Shima-2009,
  title={The super-droplet method for the numerical simulation of clouds and precipitation: 
         A particle-based and probabilistic microphysics model coupled with a non-hydrostatic model},
  author={Shima, Shin-ichiro and Kusano, Kanya and Kawano, Akio and Sugiyama, Tooru and Kawahara, Shintaro},
  journal=QJ,
  volume={135},
  number={642},
  pages={1307--1320},
  year={2009},
  publisher={Wiley Online Library}
}

@article{Seifert-2019,
author = {Seifert, Axel and Leinonen, Jussi and Siewert, Christoph and Kneifel, Stefan},
title = {The Geometry of Rimed Aggregate Snowflakes: A Modeling Study},
journal = JAMES,
volume = {11},
number = {3},
pages = {712-731},
doi = {10.1029/2018MS001519},
year = {2019}
}

@article{Brdar-Seifert-2018,
author = {Brdar, S. and Seifert, A.},
title = {McSnow: A Monte-Carlo Particle Model for Riming and Aggregation of Ice Particles 
         in a Multidimensional Microphysical Phase Space},
journal = JAMES,
volume = {10},
number = {1},
pages = {187-206},
doi = {10.1002/2017MS001167},
year = {2018}
}

@article{Libbrecht-2019,
  title={Snow Crystals},
  author={Libbrecht, Kenneth G},
  journal={arXiv preprint arXiv:1910.06389},
  doi = {10.48550/arXiv.1910.06389},
  year={2019}
}

@article{Westbrook-2004a,
  title={Theory of growth by differential sedimentation, with application to snowflake formation},
  author={Westbrook, CD and Ball, RC and Field, PR and Heymsfield, Andrew J},
  journal={Physical Review E},
  volume={70},
  number={2},
  pages={021403},
  year={2004},
  publisher={APS}
}

@article{Westbrook-2004b,
author = {Westbrook, C. D. and Ball, R. C. and Field, P. R. and Heymsfield, A. J.},
title = {Universality in snowflake aggregation},
journal = {Geophysical Research Letters},
volume = {31},
number = {15},
pages = {},
doi = {10.1029/2004GL020363},
year = {2004}
}

@article{Mitchell-1996,
  title={Use of mass-and area-dimensional power laws for determining precipitation particle terminal velocities},
  author={Mitchell, David L},
  journal=JAS,
  volume={53},
  number={12},
  pages={1710--1723},
  doi = "10.1175/1520-0469(1996)053<1710:UOMAAD>2.0.CO;2",
  year={1996}
}

@article{Mitchell-1990,
  title={Mass-dimensional relationships for ice particles and the influence of riming on snowfall rates},
  author={Mitchell, David L and Zhang, Renyi and Pitter, Richard L},
  journal=JAM,
  volume={29},
  number={2},
  pages={153--163},
  year={1990}
}

@Article{Leinonen-2021,
AUTHOR = {Leinonen, J. and Grazioli, J. and Berne, A.},
TITLE = {Reconstruction of the mass and geometry of snowfall particles from multi-angle snowflake camera (MASC) images},
JOURNAL = {Atmospheric Measurement Techniques},
VOLUME = {14},
YEAR = {2021},
NUMBER = {10},
PAGES = {6851--6866},
URL = {https://amt.copernicus.org/articles/14/6851/2021/},
DOI = {10.5194/amt-14-6851-2021}
}

@article{Locatelli-Hobbs-1974,
author = {Locatelli, John D. and Hobbs, Peter V.},
title = {Fall speeds and masses of solid precipitation particles},
journal = JGR,
volume = {79},
number = {15},
pages = {2185-2197},
doi = {10.1029/JC079i015p02185},
year = {1974}
}

@article{Morrison-2020,
  title={Confronting the challenge of modeling cloud and precipitation microphysics},
  author={Morrison, Hugh and van Lier-Walqui, Marcus and Fridlind, Ann M and Grabowski, Wojciech W
          and Harrington, Jerry Y and Hoose, Corinna and Korolev, Alexei and Kumjian, Matthew R and Milbrandt, Jason A
	  and Pawlowska, Hanna and others},
  journal=JAMES,
  volume={12},
  number={8},
  pages={e2019MS001689},
  year={2020},
  publisher={Wiley Online Library}
}

@article{Shima-2020,
  title={Predicting the morphology of ice particles in deep convection using the super-droplet method:
         development and evaluation of SCALE-SDM 0.2. 5-2.2. 0,-2.2. 1, and-2.2. 2},
  author={Shima, Shin-ichiro and Sato, Yousuke and Hashimoto, Akihiro and Misumi, Ryohei},
  journal=GMD,
  volume={13},
  number={9},
  pages={4107--4157},
  year={2020},
  doi = {10.5194/gmd-13-4107-2020},
  publisher={Copernicus Publications G{\"o}ttingen, Germany}
}

@article{Welss-2024,
  title={Explicit habit-prediction in the Lagrangian super-particle ice microphysics model McSnow},
  author={Welss, Jan-Niklas and Siewert, C and Seifert, A},
  journal=JAMES,
  volume={16},
  number={4},
  pages={e2023MS003805},
  year={2024},
  publisher={Wiley Online Library}
}

@article{Gillespie-1975,
      author = "Daniel T.  Gillespie",
      title = "An Exact Method for Numerically Simulating the Stochastic Coalescence Process in a Cloud",
      journal = JAS,
      year = "1975",
      publisher = "American Meteorological Society",
      address = "Boston MA, USA",
      volume = "32",
      number = "10",
      pages=      "1977 - 1989"
}

@article{Leinonen-Moisseev-2015,
  title={What do triple-frequency radar signatures reveal about aggregate snowflakes?},
  author={Leinonen, J and Moisseev, D},
  journal=JGR,
  volume={120},
  number={1},
  pages={229--239},
  year={2015},
  publisher={Wiley Online Library}
}

@article{Leinonen-Szyrmer-2015,
author = {Leinonen, Jussi and Szyrmer, Wanda},
title = {Radar signatures of snowflake riming: A modeling study},
journal = {Earth and Space Science},
volume = {2},
number = {8},
pages = {346-358},
doi = {10.1002/2015EA000102},
year = {2015}
}

@article{Ori-2021,
  title={snowScatt 1.0: consistent model of microphysical and scattering properties of rimed and
         unrimed snowflakes based on the self-similar Rayleigh--Gans approximation},
  author={Ori, Davide and von Terzi, Leonie and Karrer, Markus and Kneifel, Stefan},
  journal=GMD,
  volume={14},
  number={3},
  pages={1511--1531},
  year={2021},
  publisher={Copernicus Publications G{\"o}ttingen, Germany}
}

@article{Karrer-2020,
author = {Karrer, M. and Seifert, A. and Siewert, C. and Ori, D. and von Lerber, A. and Kneifel, S.},
title = {Ice Particle Properties Inferred From Aggregation Modelling},
journal = JAMES,
volume = {12},
number = {8},
pages = {e2020MS002066},
doi = {10.1029/2020MS002066},
year = {2020}
}

@article{Hales-2024,
      title={The Formal Proof of the Kepler Conjecture: a critical retrospective}, 
      author={Thomas Hales},
      year={2024},
      journal={arXiv},
      eprint={2402.08032},
      archivePrefix={arXiv},
      primaryClass={math.MG},
      url={https://arxiv.org/abs/2402.08032}, 
}

@article{Ball-2011,
  title={In retrospect: On the six-cornered snowflake},
  author={Ball, Philip},
  journal = {Nature},
  year={2011},
  volume = {480},
  number = {450},
  doi={10.1038/480455a},
  publisher={Nature Publishing Group UK London}
}

@article{Boehm-1992a,
title = "A general hydrodynamic theory for mixed-phase microphysics. Part I: drag and fall speed of hydrometeors",
journal = "Atmos. Res.",
volume = "27",
number = "4",
pages = "253 - 274",
year = "1992",
issn = "0169-8095",
doi = "10.1016/0169-8095(92)90035-9",
author = {Böhm, Johannes P.},
}

@article{Boehm-1992b,
title = "A general hydrodynamic theory for mixed-phase microphysics. Part II: collision kernels for coalescence",
journal = "Atmos. Res.",
volume = "27",
number = "4",
pages = "275 - 290",
year = "1992",
issn = "0169-8095",
doi = "10.1016/0169-8095(92)90036-A",
author = {Böhm, Johannes P.},
}

@article{Boehm-1992c,
title = "A general hydrodynamic theory for mixed-phase microphysics. Part III: Riming and aggregation",
journal = "Atmos. Res.",
volume = "28",
number = "2",
pages = "103 - 123",
year = "1992",
issn = "0169-8095",
doi = "10.1016/0169-8095(92)90023-4",
author = {Böhm, Johannes P.},
}

@article{Shaw-2003,
  title={Particle-turbulence interactions in atmospheric clouds},
  author={Shaw, Raymond A},
  journal={Annual Review of Fluid Mechanics},
  volume={35},
  number={1},
  pages={183--227},
  year={2003},
  publisher={Annual Reviews, Palo Alto, CA, USA}
}

@article{Grabowski-Wang-2013,
  title={Growth of cloud droplets in a turbulent environment},
  author={Grabowski, Wojciech W and Wang, Lian-Ping},
  journal={Annual review of fluid mechanics},
  volume={45},
  number={1},
  pages={293--324},
  year={2013},
  publisher={Annual Reviews}
}

@article{Onishi-Seifert-2016,
  title={Reynolds-number dependence of turbulence enhancement on collision growth},
  author={Onishi, Ryo and Seifert, Axel},
  journal=ACP,
  volume={16},
  number={19},
  pages={12441--12455},
  year={2016},
  doi = {10.5194/acp-16-12441-2016}
}

@article{Kobschall-2023,
  title={Geometric descriptors for the prediction of snowflake drag},
  author={K{\"o}bschall, Kilian and Breitenbach, Jan and Roisman, Ilia V and Tropea, Cameron and Hussong, Jeanette},
  journal={Exp. Fluids},
  volume={64},
  number={1},
  pages={4},
  year={2023},
  doi={10.1007/s00348-022-03539-x},
  publisher={Springer}
}

@article{Dunnavan-2019,
  author    = {Dunnavan, Emma L. and Jiang, Zhien and Harrington, Jerry Y. and Verlinde, Johannes and Fitch, Kevin and Garrett, Timothy J.},
  title     = {The Shape and Density Evolution of Snow Aggregates},
  journal   = JAS,
  year      = {2019},
  volume    = {76},
  number    = {12},
  pages     = {3919--3940},
  doi       = {10.1175/JAS-D-19-0066.1},
  url       = {https://doi.org/10.1175/JAS-D-19-0066.1}
}

@article{Przybylo-2022,
  author    = {Przybylo, Vanessa M. and Sulia, Kara J. and Lebo, Zachary J. and Schmitt, Carl G.},
  title     = {The Ice Particle and Aggregate Simulator (IPAS). Part II: Analysis of a Database of Theoretical
               Aggregates for Microphysical Parameterization},
  journal   = JAS,
  year      = {2022},
  volume    = {79},
  number    = {6},
  pages     = {1633--1649},
  doi       = {10.1175/JAS-D-21-0179.1}
}

@article {Telford-1955,
      author = "J. W.  Telford",
      title = {A New Aspect of Coalescence Theory},
      journal = JAS,
      year = "1955",
      publisher = "American Meteorological Society",
      address = "Boston MA, USA",
      volume = "12",
      number = "5",
      pages= "436 - 444"
 }

@article{Morrison-2024,
  title={Impacts of stochastic coalescence variability on warm rain initiation using Lagrangian microphysics in box and large-eddy simulations},
  author={Morrison, Hugh and Chandrakar, Kamal Kant and Shima, Shin-Ichiro and Dziekan, Piotr and Grabowski, Wojciech W},
  journal=JAS,
  volume={81},
  number={6},
  pages={1067--1093},
  year={2024},
  publisher={American Meteorological Society}
}

@article {Chandrakar-2021,
      author = "Kamal Kant Chandrakar and Wojciech W. Grabowski and Hugh Morrison and George H. Bryan",
      title = "Impact of Entrainment Mixing and Turbulent Fluctuations on Droplet Size Distributions in a Cumulus Cloud:
                An Investigation Using Lagrangian Microphysics with a Subgrid-Scale Model",
      journal = JAS,
      year = "2021",
      publisher = "American Meteorological Society",
      address = "Boston MA, USA",
      volume = "78",
      number = "9",
      doi = "10.1175/JAS-D-20-0281.1",
      pages=      "2983 - 3005",
}

@article{Dunnavan-2021,
  title={How snow aggregate ellipsoid shape and orientation variability affects fall speed and self-aggregation rates},
  author={Dunnavan, Edwin L},
  journal=JAS,
  volume={78},
  number={1},
  pages={51--73},
  year={2021},
  doi={10.1175/JAS-D-20-0128.1}  
}

@article{Chabrier-2003,
  author = {Chabrier, Gilles},
  title = {Galactic Stellar and Substellar Initial Mass Function},
  journal = {Publications of the Astronomical Society of the Pacific},
  year = {2003},
  volume = {115},
  pages = {763--795},
  doi = {10.1086/376392},
}

@article{Basu-Jones-2004,
  author = {Basu, Shantanu and Jones, Christopher E.},
  title = {A Modified Lognormal Power-law Distribution for the Stellar Initial Mass Function},
  journal = {Monthly Notices of the Royal Astronomical Society Letters},
  year = {2004},
  volume = {347},
  pages = {L47--L51},
  doi = {10.1111/j.1365-2966.2004.07341.x},
}

@article{Chellini-Kneifel-2024,
author = {Chellini, Giovanni and Kneifel, Stefan},
title = {Turbulence as a Key Driver of Ice Aggregation and Riming in Arctic Low-Level Mixed-Phase Clouds, Revealed by Long-Term Cloud Radar Observations},
journal = GRL,
volume = {51},
number = {6},
pages = {e2023GL106599},
doi = {10.1029/2023GL106599},
year = {2024}
}

\end{document}